\documentclass[12pt]{article}
\usepackage{cite}
\usepackage{epsf}
\def\beq{\begin{equation}}   \def\eeq{\end{equation}}

\begin{document}

\begin{titlepage}
\renewcommand{\thefootnote}{\fnsymbol{footnote}}

\begin{flushright}
TPI-MINN-98/20-T\\
UMN-TH-1724/98\\
hep-th/9810068\\

\end{flushright}

\vspace{0.3cm}

\begin{center}
\baselineskip25pt

{\Large\bf Anomaly and  Quantum Corrections to
Solitons in Two-Dimensional Theories with Minimal Supersymmetry}

\end{center}

\vspace{0.3cm}

\begin{center}
\baselineskip12pt

\def\thefootnote{\fnsymbol{footnote}}

{\large M. Shifman, A. Vainshtein}

\vspace{0.2cm}
{\it
  Theoretical Physics Institute, University of Minnesota,
Minneapolis, MN 55455
\\[0.2cm]}
and \\[0.2cm]
{\large M. Voloshin}\\[0.2cm]
{\it
  Theoretical Physics Institute, University of Minnesota,
Minneapolis, MN 55455
\\
and \\
Institute of Experimental and Theoretical Physics, Moscow 117259}
\vspace{1.5cm}

{\large\bf Abstract} \vspace*{.25cm}
\end{center}

We reexamine the issue of the soliton mass in two-dimensional models with
${\cal N} =1$ supersymmetry. The superalgebra has a  central extension,
and at the classical level the soliton solution preserves 1/2 of
supersymmetry which is equivalent to BPS saturation. We prove that the
property of BPS saturation, i.e. the equality of the soliton mass to the
central charge, remains intact at the quantum level in all orders of the
weak coupling expansion. Our key finding is an anomaly in the expression
for the central charge. The classical central charge, equal to the jump
of the superpotential, is amended by an anomalous term proportional to
the second derivative of the superpotential. The anomaly is established
by various methods in  explicit one-loop calculations. We argue that
this one-loop result is not affected by higher orders.  We discuss in
detail  how the impact of the boundary conditions can be untangled from
the soliton mass calculation. In particular, the soliton profile and the
energy distribution are found at one loop. A ``supersymmetry" in the
soliton mass calculations in the non-supersymmetric models is observed.
\vspace{1cm}

\noindent
Phys. Rev. D, to appear
\end{titlepage}

\tableofcontents

\newpage

\section{Introduction}
\label{sec:intro}
\setcounter{equation}{0}

The issue of quantum corrections to the soliton mass in two-dimensional
models with ${\cal N} =1$ supersymmetry has a long and dramatic history.
As was noted in \mbox{Ref.\ \cite{WO}},  in the models with central
extensions topologically stable solitons can be Bogomol'nyi-Prasd-Sommerfeld (BPS)
saturated. Such solitons are annihilated by the action of 1/2 of the supergenerators,
so that 1/2 of the supersymmetry (SUSY) is preserved. The remaining  1/2 of
the supergenerators act non-trivially.  The mass of the BPS saturated
solitons must be  equal to the central charge. The simplest
two-dimensional model with the {\em minimal} superalgebra, admitting
solitons, was considered in \cite{DADI}.  This is the Wess-Zumino model
with one real scalar field and one two-component real spinor (to be
referred to as the {\em minimal model} hereafter). It was argued
\cite{DADI} that, due to a residual supersymmetry, the mass of the
soliton calculated at the classical level remains intact at  the
one-loop level. A few years later it was noted \cite{KR} that the
non-renormalization theorem \cite{DADI} cannot possibly be correct,
since the classical soliton mass is proportional to $m^3/\lambda^2$
(where $m$ and $\lambda$ are the bare mass parameter and coupling
constant, respectively), and the physical mass of the scalar field gets
a logarithmically infinite renormalization. Since the soliton mass is an
observable physical parameter, it must stay finite in the limit $M_{\rm
uv}\to\infty$, where $M_{\rm uv}$ is the ultraviolet cut off. This
implies, in turn, that the quantum corrections cannot vanish -- they
``dress" $m$ in the classical expression, converting  the bare mass
parameter into the renormalized one.  The one-loop renormalization of
the soliton mass was first calculated in \cite{KR}. Technically the
emergence of the one-loop correction was attributed to a ``difference in
the density of states in continuum in the boson and fermion operators in
the soliton background field".  The subsequent work \cite{IM} dealt with
the renormalization of the central charge, with the conclusion that the
central charge is renormalized in just the same way as the soliton mass,
so that the saturation condition is not violated.

Since then a number of one-loop calculations have been carried out
\cite{AHV,JFS,SR,AU,AU1,HY,CM1,CM,RN,NSNR,Jaffe}. The results reported
and the  conclusion of saturation/non-saturation oscillate with time,
with no
sign of convergence. Although all authors clearly agree that the
logarithmically divergent term corresponds to the renormalization of
$m$, the finite term comes out differently,  varying from work to work,
sometimes even in the successive works of the same authors, e.g.
\cite{RN,NSNR} or \cite{AU,AU1}. According to a recent publication
\cite{NSNR}, the BPS saturation is violated at one loop. This assertion
reverses the earlier trend \cite{KR,HY,CM1}, according to which the
soliton  in the minimal model is BPS saturated: the central charge and
the soliton  mass are renormalized in a concerted way.

The problem becomes obscure partly due to the fact that some authors
(e.g.  \cite{KR})  use  non-supersymmetric boundary conditions. When
non-supersymmetric  boundary conditions are imposed, explicit
supersymmetry  is lost. Gone  with it is the cancellation of the boson
and fermion loops inherent to  supersymmetry. Then one has to consider
separately the boson and   fermion contributions -- each of them
individually diverges. Extracting  finite terms in the difference of two
diverging contributions complicates  the analysis. This approach -- the
separate treatment of the boson and fermion corrections in a
SUSY-insensitive manner -- prevails in the recent works. 
Although such approach is fully legitimate, it does
not take advantage of SUSY.

We reexamine the issue of the quantum corrections to solitons in
two-dimensional models with  minimal supersymmetry. Our strategy is
to take  maximal advantage of  supersymmetry, at every step. One of our
main goals is establishing a relation with other similar problems where
1/2 of SUSY is preserved in the background field: the instanton
calculations in SUSY gluodynamics \cite{NSVZbeta} and the BPS-saturated
domain walls \cite{DS,CS}. In both cases the quantum corrections totally
cancel, bosons against fermions. Up to a point, a similar cancellation takes
place in the two-dimensional models too. If it were not for boundary
effects, the classical expression for the soliton mass  would not be
corrected at the quantum level. The boundary effects, which play no role
in SUSY instantons  or the BPS-saturated domain walls, are responsible
for the soliton mass renormalization in the two-dimensional models with
minimal supersymmetry. The divergence between the two- and
four-dimensional problems lies here.  A non-renormalization theorem
survives in the form of the statement that the soliton mass is equal to the
central charge, to {\em all orders}.  We present a  general argument for
why the BPS saturation cannot be violated by quantum corrections in the
weak coupling regime. In four dimensions the phenomenon known as
multiplet shortening protects the BPS states from becoming nonsaturated
at higher orders. In two dimensions the multiplet shortening does not
occur, literally speaking. However, the multiplet shortening {\em per se}
is {\em not} the reason lying behind the protected nature of the BPS
states, but, rather, a signature. The genuine reason is the existence of an
invariant subalgebra of the superalgebra, i.e. the existence of a
supercharge $Q$ possessing the following properties: (i) $Q^2$ is
nontrivial; (ii) $Q$ annihilates all states from the given multiplet.
Although the multiplet shortening does entail the existence of such a
subalgebra, the inverse is not necessarily true. In fact, the
two-dimensional models with minimal supersymmetry present an
example of the problem where the invariant subalgebra of the required
type exists without the multiplet shortening. This is sufficient for
establishing the theorem ``soliton mass $=$ central charge", order by
order.

The next question to address is: ``what is  the central charge in the
theories with minimal supersymmetry?" This is a key element of our
analysis. As we will show below, much of the confusion in the current
literature is due to a tacit assumption that the central charge
${\cal Z}$ is equal to the difference between the values of the 
superpotential ${\cal W}$ at spatial infinities,
\begin{equation}
{\cal Z} = {\cal  W}[\phi (z=\infty )]-{\cal W}[\phi (z=-\infty )]\, .
\label{zfirst}
\end{equation}
This expression was obtained long ago \cite{WO}.  At the classical
level it is a valid expression. It was implied in the previous studies
that the only change at the quantum level is the substitution of the
classical superpotential ${\cal W}$ in Eq.\ (\ref{zfirst} ) by its
operator form. We show that the central charge is additionally
modified by a {\it quantum anomaly}, 
\beq 
{\cal W}\longrightarrow
{\cal W} +\frac{{\cal W}'' }{4\pi}\, ,
\label{qa}
\eeq
i.e. the anomalous term ${\cal W}''$ should be added in the expression 
for the central charge.

We have obtained this anomaly by four different methods: (i) using
dimensional regularization; (ii) regularization by higher derivatives;
(iii) from an effective superpotential; (iv) from the models with the
extended (${\cal N}=2$) supersymmetry deformed by soft terms breaking of
${\cal N}=2$ down to ${\cal N}=1$.  If the first two methods are quite
straightforward, the third and the fourth require comment. As in many
other similar problems, it is convenient to  calculate the soliton mass
starting from  an effective Lagrangian (superpotential), including the
loop effects by construction, rather than from the bare Lagrangian. It
is not difficult to obtain the effective superpotential at one-loop
level, which results, in turn, in a certain prediction for the soliton
mass. This prediction can be compared with the result  obtained through
the central charge. The two calculations agree with each other provided
one accounts for the anomaly in Eq. (\ref{qa}).

An alternative approach is based on the models with  extended
supersymmetry, ${\cal N}=2$.  The  ${\cal N}=2$ theories in two
dimensions  can be obtained by dimensional reduction of the Wess-Zumino
model \cite{WZ},  from four to two dimensions.  The superpotential is
not renormalized \cite{GRS}, and so neither is  the soliton mass. The
boundary effects  --  the prime culprits for  the soliton mass
renormalization in the minimal model -- vanish in the ${\cal N}=2$ case.
We then deform the ${\cal N}=2$ model by introducing terms which softly
break ${\cal N}=2$ SUSY down to ${\cal N}=1$. The emerging dynamics is
quite rich and interesting. On the technical side, the softly broken
${\cal N}=2$ theories provide a possible regularization procedure for
${\cal N}=1$ theories. Applying this procedure we recover 
Eq.\ (\ref {qa}) again, for the fourth time.

Although the fact that the central charge does have anomalies is very
well known in the four-dimensional theories (e.g. \cite{CS}), the
existence of the anomaly in two-dimensional theories with minimal
supersymmetry was overlooked in all previous publications. Needless to
say that the additional term proportional to ${\cal W}'' $ in Eq.
(\ref{qa}) gives a non-vanishing contribution to the central charge. In
conjunction with the statement of BPS saturation, this contribution
translates into a (finite) extra term in the soliton mass. If the BPS
saturation {\em per se} is established on the basis of comparison
between two distinct calculations -- of the soliton mass, on the one
hand, and of the central charge on the other -- the omission  of the
anomalous term  leads to an apparent paradox. The general theorem
forbidding the loss of BPS saturation in quantum loops in the weak
coupling expansion seems to be violated.

Since the quantum corrections are associated with boundary effects,
their determination reduces to a calculation  in the ``empty" vacuum,
which can be trivially performed.  At the same time, the standard
calculation of the soliton mass becomes sensitive to the choice of
boundary conditions. This is another subtle point which was not properly
treated in the literature.  It was believed that the soliton mass
depends on the boundary conditions, and that particular choices of the
boundary conditions are ``more correct" than others. It is clear that
physically the soliton mass is saturated in the domain near its center,
and must not depend on the boundary conditions set at a large distance
from the soliton center.

Although the boundary conditions do not affect the physical soliton
mass, some choices are practically more convenient. In fact,
inappropriate choices contaminate the field near the boundary and may
shift the total energy in the box  by a quantity which is of the same 
order as the ``genuine" one-loop quantum correction to the kink mass. 
If the boundary domain is not
treated properly, the resulting calculation yields the mass of a
contaminated soliton, rather than the physical mass of the ground state
in the sector with unit topological charge (i.e. the true soliton). Much
of the  controversy existing in the literature is due to this
contamination.  We explicitly demonstrate how these distortions appear
in the boundary domain, and show how the boundary conditions can be
untangled from the soliton mass calculations. One of the possibilities
to control  the  contamination is to consider the  energy-density
distribution in the soliton at one loop. This is done in 
Sec.\ \ref{sec:profile}. In addition,  we perform a calculation of the
one-loop correction to the soliton profile.

The last issue addressed in this work is the soliton mass calculation in
the non-supersymmetric sine-Gordon model. Surprisingly, this problem is
naturally formulated in supersymmetric terms too, although the original
Lagrangian  has no explicit supersymmetry. The role of the SUSY partners
is played in the case at hand by the sectors with
the zero and unit topological charges, respectively. This approach
yields a number of useful results. In particular, closed analytical
expressions for the boson and fermion Green functions in  the soliton
background,  known previously \cite{mv,bs}, are given a  transparent
interpretation.

The outline of the paper is  as follows. In Sec.\ \ref{sec:minimal}  we
briefly review the formulation of the problem, and general aspects of
two-dimensional theories with minimal supersymmetry. Most of the results
presented here are not new. Since they are scattered in the literature
we find it convenient to collect them in one place, organizing a common
perspective. We briefly review the superfield formalism in two
dimensions and describe the two models with minimal supersymmetry we
will  deal with: the polynomial model and the sine-Gordon model. Then
the kink solutions are discussed, and the notion of the kink superfield
is introduced. A  residual supersymmetry which annihilates the  kink is
identified. At the classical level, the corresponding SUSY generator
acts on the kink trivially. It is shown that this property cannot be
lost in perturbation theory. In Sec.\ \ref{ssec:WI} we discuss Witten's
index in two-dimensional models with minimal supersymmetry.

Section  \ref{regularization} is devoted to derivations of the central
charge. The focus is on the anomalous part of the central charge, which
is obtained in a variety of ways. We formulate a convenient (and
universal) regularization procedure based on higher derivatives, which
regularizes all ultraviolet divergences. We also explain how the same
anomaly emerges as an infrared effect, through the level flow. This is a
standard situation -- the anomaly has two faces: ultraviolet and
infrared (see e.g. \cite{MiS}). In Sec.\ \ref{ssec:local} a local version
of the BPS saturation condition is explored.
In Sec.\ \ref{sec:onemass} the one-loop corrections to the kink mass
are calculated, both in general form and in the two models under
consideration, the polynomial model and sine-Gordon. We present two
distinct calculations.
Section \ref{sec:profile} deals with the soliton profile at one loop.
The energy-density distribution in the kink is also presented here.
The results are based on the explicit expressions for Green's function
in the soliton background.
In Sec.\ \ref{sec:recurrency} the calculations of the soliton masses in
non-supersymmetric polynomial and sine-Gordon models are reformulated in
terms of a supersymmetry which -- surprisingly -- can be introduced in
these problems. 
Section \ref{2loops} presents the calculation of the soliton mass in the
supersymmetric sine-Gordon model at two loops. Finally, 
Sec.\ \ref{sec:extended} is devoted to extended supersymmetry and ${\cal N}
=1$ deformations of the ${\cal N} =2$ model. Section \ref{sec:disc}
summarizes our findings.

\section{The minimal SUSY in two dimensions}
\label{sec:minimal}
\setcounter{equation}{0}

\subsection{Superspace for ${\cal N}$=1 SUSY models}
\label{ssec:super}

 The two-dimensional space
$x^\mu = \{t,z\}$ can be promoted to  superspace by adding
a two-component {\em real} Grassmann variable $\theta_\alpha =\{
\theta_1, \theta_2\}$. The coordinate transformations
\beq
\theta_\alpha \to \theta_\alpha +\varepsilon_\alpha\, , \quad x^\mu
\to
x^\mu  - i\bar\theta\gamma^\mu \varepsilon\,
\label{str2dim}
\eeq
add SUSY to the translational and Lorentz invariances.  A
convenient representation for the two-dimensional Majorana
$\gamma$ matrices we use throughout this paper is
\beq
\gamma^0 = \sigma_2\, , \quad \gamma^1 = i\sigma_3\, .
\eeq
The real superfield $\Phi \,(\! x,\theta ) $ has the form
\beq
\Phi\,(\!x,\theta ) = \phi \, (\! x) +\bar\theta\psi (\! x)
+\frac{1}{2}\bar\theta\theta F (\! x)\, ,
\label{n1super}
\eeq
where
$\bar\theta = \theta\gamma^0$, and $\theta , \psi$ are real
two-component spinors.
The superspace transformations (\ref{str2dim})  generate the
following
SUSY transformations:
\beq
\delta\phi = \bar\varepsilon \psi\, ,\quad
\delta\psi = -i \, \partial_\mu \phi \,\gamma^\mu\varepsilon
+ F \varepsilon\, , \quad
\delta F =-i \bar\varepsilon \gamma^\mu \partial_\mu \psi \, .
\label{susytr}
\eeq
In 1+1 dimensions it is a minimal, ${\cal N}=1$, supersymmetry.

The action of a model invariant under
the  transformations (\ref{susytr})  is
\beq
S =i\int {\rm d}^2\theta\, {\rm d}^2x \left\{ \frac{1}{4}\bar D_\alpha
\Phi
D_\alpha\Phi +{\cal W}(\Phi )\right\}
\label{minac}
\eeq
where ${\cal W}(\Phi )$ will be referred to as the superpotential,
keeping in mind a parallel with the four-dimensional Wess-Zumino
model, although in the case at hand the superpotential term is
the integral over the full superspace, and is not chiral.
Moreover, ${\rm d}^2 x = {\rm d}t \, {\rm d}z$;
the spinorial derivatives are defined as follows
\beq
D_\alpha = \frac{\partial}{\partial\bar\theta_\alpha}
- i \,(\!\gamma^\mu\theta)_\alpha\partial_\mu\, , \quad
{\bar D}_\alpha = -\frac{\partial}{\partial\theta_\alpha}
+ i \,(\!\bar\theta\gamma^\mu )_\alpha\partial_\mu\, ,
\eeq
so that
$$
\{D_\alpha\bar D_\beta\} = 2i\,(\!\gamma^\mu
)_{\alpha\beta}\partial_\mu\, .
$$
In components
the Lagrangian takes the form
\beq
{\cal L} = \frac{1}{2}\left( \partial_\mu\phi \,\partial^\mu\phi
+\bar\psi\, i\! \!\not\!\partial \psi +F^2 \right) + {\cal W}'(\phi)F
-\frac{1}{2}{\cal W}''(\phi)\bar\psi\psi\, .
\label{minlag}
\eeq

In two dimensions any superpotential function ${\cal W}$ leads to a
renormalizable field theory and is thus allowed. Many of the results
presented below are independent of the particular choice of
${\cal W}$. For our purposes it is generally sufficient to limit
ourselves to odd
functions of $\Phi$, i.e. ${\cal W}(\Phi ) $= $- {\cal W}(-\Phi )$
(\mbox{Sec.\ \ref{sec:extended}} is  exception). We require the
superpotential ${\cal W}(\phi)$ to have more than one extremum, i.e. the
equation ${\rm d}\,{\cal W}/{\rm d}\phi = 0$ must have at least two
solutions. Then one  deals with several degenerate vacua. The field
configurations interpolating between distinct vacua at the  spatial
infinities \mbox{($z\to\pm\infty$)} are topologically stable solitons (kinks).

In certain instances we find it illustrative to consider specific
examples. Two choices are of most practical interest: the polynomial
superpotential, and the minimal supergeneralization of the sine-Gordon
model. In the first case the superpotential function ${\cal W}$ is
parameterized as follows:
\beq
{\cal W}(\Phi) = \frac{m^2}{4\,\lambda} \,\Phi - \frac{\lambda}{3}
\,\Phi^3\, ,
\label{spot}
\eeq
where the parameters $m$ and $\lambda$ are  real numbers.
In the second case
\beq
{\cal W}(\Phi) = m v^2 \,
\sin\frac{\Phi}{v}\, .
\label{spotsg}
\eeq
The first model will be referred to as SPM, the second
as SSG.

\subsection{Multiple vacua and kinks}
\label{ssec:kinks}

The SPM model  has  two vacuum states
corresponding to
 $\phi^\pm_* =\pm m /(2\lambda)$. A  kink solution
interpolates between  $\phi^-_* =- m /2\lambda$ at $z\to -\infty$
and
$\phi^+_* =  m  /2\lambda$ at $z\to \infty$, while an antikink
between
$\phi^+_* = m /2\lambda$ and
$\phi^-_* = - m /2\lambda$.   The classical kink solution has the
form
\beq
\phi_{\,0} = \frac{m}{2\lambda}\,  \tanh \frac{mz}{2}\, .
\label{th}
\eeq

In the sine-Gordon model there are infinitely many vacua; they lie at
\beq
\phi^{\,k}_* = v \left( \frac{\pi}{2} + k \pi\right)
\eeq
where $k$ is an integer, either positive or negative. Correspondingly,
there exist solitons connecting any pair of vacua. In this case
we will limit ourselves to consideration of the ``elementary" solitons
connecting the adjacent vacua, e.g.
$
\phi^{\;0,-1}_*= \pm \pi v/2
$,
\beq
\phi_{\,0} = v\arcsin \left[ \, \tanh
(mz)\right]\, .
\label{thsg}
\eeq

Both solutions (\ref{th}) and (\ref{thsg}) are BPS saturated, i.e. they
satisfy the first order differential equation,
\beq
\frac{ {\rm d} \phi_{\,0}(z)}{{\rm d} z} =  - F (\phi_{\,0})= {\cal
W}^\prime
(\phi_{\,0}) \, .
\label{solution}
\eeq
In terms of superfields  the BPS saturation means that
 a  background field $\Phi_0$ is of the form
\beq
\Phi_0 (x,\theta ) = f \left( z- \frac{1}{2}\bar\theta\theta\right)\, ,
\label{clbf}
\eeq
where $f$ is some function. In other words, instead of the generic
superfield
$\Phi (x,\theta ) $ depending on two coordinates $x^\mu$ and two
Grassmann coordinates $\theta_\alpha$ we deal with a background
(super)field depending only on one specific combination, $z-
\bar\theta\theta
/2$. In components \mbox{Eq.\ (\ref{clbf})} implies that
\beq
 \phi_{\,0}= f(z), \quad F_{\,0}= -\frac{ {\rm d} f(z)}{{\rm d} z}\; .
\label{scbf}
\eeq
The soliton solutions  (\ref{th}) and (\ref{thsg}) belong to this class,
with a special choice of the function $f(z)$.

A special, topological nature of the background field (\ref{clbf})
is seen from the following property:   the total energy
is defined by the asymptotics at $z\to \pm \infty$,
\beq
 -\int {\cal L}[\Phi_0]dz =\Delta {\cal W}\equiv {\cal
W}[\phi (z=\infty )]-
 {\cal W}[\phi (z=-\infty )]\, ,
\label{fsolmas}
\eeq
irrespective of the particular choice of the function $f$.
Substituting the asymptotic values of $\phi$ at $\pm\infty$
in \mbox{Eq.\ (\ref{fsolmas})} we find the soliton mass
\beq
M_0 = \left\{ \begin{array}{l}
\mbox{\large $\frac{m^3}{6\,\lambda^2}
$}
\quad \qquad\
~\mbox{(SPM)}
\\[0.2cm]
\mbox{$2\, m v^2$}
\quad \qquad
\mbox{(SSG)}
\end{array}\right.
\label{csm}
\eeq
where the subscript 0 emphasizes that this result is obtained at the
classical level.

 For {\em any} background field of the type~(\ref{clbf})
half of the supersymmetry is preserved. Indeed, let us consider
supertransformations~(\ref{susytr}) with the parameter
$\varepsilon_2=0$,
\begin{equation}
\delta\theta_1=\varepsilon_1\,,\quad  \delta\theta_2=0\,, \quad
\delta t=-i \theta_1 \varepsilon_1\,,\quad \delta z=i \theta_2
\varepsilon_1\,.
\label{susytr1}
\end{equation}
This  transformation   leaves the interval
\beq
z_{\,\rm inv}\equiv z- \frac{1}{2}\bar\theta\theta=z+i\theta_1
\theta_2
\label{invint}
\eeq
invariant.
Consequently,  backgrounds of the form (\ref{clbf})
are not changed.  It means that quantization  preserves  the
invariance under the following transformations of the fields,
\begin{equation}
\delta\left[ \begin{array}{c}
\phi\\ -i \psi_2 \end{array} \right]=\varepsilon_1
\left[ \begin{array}{c}
 -i \psi_2 \\ -i \dot\phi \end{array} \right]\,,\quad\quad
\delta\left[ \begin{array}{c}
 \psi_1\\ \partial_z \phi +F \end{array} \right]=\varepsilon_1
\left[ \begin{array}{c}
 \partial_z \phi +F \\ -i\dot\psi_1 \end{array} \right]\,.
\label{halfsusytr}
\end{equation}
This form shows the  multiplet structure of the surviving supersymmetry.
An important consequence of the invariance~(\ref{halfsusytr}) is that
the fermion-boson degeneracy is not lost for quantum excitations
although a part of SUSY  is broken in the given  fixed background.

The part of supersymmetry which is spontaneously broken by
the soliton configuration corresponds to the transformations produced by
$\varepsilon_2$,
\begin{equation}
\delta\theta_1=0\,,\quad  \delta\theta_2=\varepsilon_2\,, \quad
\delta t=-i \theta_2 \varepsilon_2\,,\quad \delta z=i \theta_1
\varepsilon_2\,.
\label{susytr2}
\end{equation}
Applied to the soliton solution it generates the fermionic field
$\psi_2$
(use \mbox{Eq.\ (\ref{susytr})} with $\varepsilon_1=0$):
\begin{equation}
\delta \psi_2= (-\partial_z \phi_0 +F_0)\, \varepsilon_2=-2\,
\frac{{\rm d}
\phi_0 (z)}{ {\rm d} z}\, \varepsilon_2\;.
\end{equation}
In addition to the above fermion zero mode 
there is also the bosonic zero mode generated by the spatial
translation.  Both zero modes can be conveniently described by the
introduction of two collective coordinates: the soliton center $z_{\,0}$
and the real Grassmann parameter $\eta $ corresponding to the
fermion zero  mode, in much the same way this was done for
instantons \cite{NSVVZ} and domain walls \cite{CS}.
To this end let us make the substitution
\beq
z_{\,\rm inv} = z +i\theta_1 \theta_2 \quad \longrightarrow \quad
\tilde z_{\,\rm inv} = z-
z_{\,0} +i\theta_1 (\theta_2-\eta)
\, .
\eeq
The kink superfield, with all collective coordinates included, is
 \begin{equation}
\Phi_0 = f\, (\tilde z_{\rm inv})\;,
\end{equation}
where the function $f$ is given by \mbox{Eq.\ (\ref{th})} for the
polynomial model and by \mbox{Eq.\ (\ref{thsg})} for the sine-Gordon
one.

The deformation of the kink superfield $\Phi_0$ under the
supertransformations~(\ref{str2dim}) is compensated by
the following transformations of the collective coordinates:
\begin{equation}
\delta z_{\,0}= i \eta \,\varepsilon_1\,,\quad \delta\eta=2
\varepsilon_2\,.
\label{collect}
\end{equation}

The parameters $z_{\,0}$ and $\eta$ can be treated as
time-dependent dynamical variables, describing the superkink.
The quantum mechanics of these coordinates
emerges upon integration over all non-zero modes
in the field-theoretical model at hand,
\begin{eqnarray}
Z = \int {\cal D}[\phi (x)]{\cal D}[\psi (x)]\exp\{ i\!\int\!{\cal
L}(\phi , \psi
){\rm d}^2 x \}= ~~~~~~\nonumber\\
\int {\cal D}[z_{\,0}(t)]{\cal D}[\eta (t) ]\exp \left\{
i\int \! {\rm d} t\left[-M +
\frac{1}{2}M\dot{z}_{\,0}^2+\frac{iM}{2}\eta\dot\eta\right]\right\}
\,,
\label{frstqu}
\end{eqnarray}
where the functional integral is performed in the sector with unit
kink number, and $M$ is the physical mass of the kink.

\subsection{Witten's index in the models with minimal
supersymmetry}
\label{ssec:WI}

Here we comment on a general aspect of the two-dimensional models
with minimal supersymmetry which is only indirectly related to
the solitons. In the model (\ref{minac}), (\ref{minlag}) with the
polynomial potential the Witten index  is zero.
This can be seen in many ways. Say, if the mass parameter $m^2$
in the superpotential (\ref{spot}) is made negative, the equation
${\rm d}{\cal W}/{\rm d}\Phi = 0$ has no real solutions.
At negative $m^2 $ the $Z_2$ symmetry
$$\phi \to -\phi\, , \quad \psi \to \gamma_5\psi
$$
 stays unbroken. This protects
$\phi$ from developing a vacuum expectation value.
Therefore, the $\psi$ field is massless, and is the Goldstino of the
spontaneously broken SUSY.  A convenient order parameter is
$\langle {\cal W}^\prime \rangle$, where
\beq
{\cal W}^\prime \equiv \frac{{\rm d} {\cal W}}{{\rm d} \Phi }=
\frac{m^2}{4\lambda} - \lambda \Phi ^2\, .
\label{doubleu}
\eeq
Since ${\cal W}' = (1/2) \bar{D}D\Phi$, the  expectation
value
$\langle {\cal W}' \rangle = 0$  in  the supersymmetric vacuum.

Alternatively, one can consider the theory with positive $m^2$ in
a finite box. If, following Witten \cite{EW}, we make the box width
$L$ small, $\lambda L\ll 1$,  and retain only the zero momentum
modes, discarding all  others, we get a quantum-mechanical system
known to have a vanishing Witten index \cite{EW}.

At positive (and large) values of $m^2$ the $Z_2$  symmetry is
spontaneously broken, so that the theory has two physically
equivalent vacua. The vanishing of the Witten index implies that  if
one of the vacua is bosonic, the other is fermionic. This is a unique
feature of the minimal model: in four-dimensional field theories and in
two-dimensional field theories with extended supersymmetry (and no
massless particles) all vacuum states are clearly of the bosonic type.

All theories with vanishing Witten index are potentially unstable
with respect to spontaneous SUSY breaking. Moreover, in a finite
volume this spontaneous breaking does occur. However, in the infinite
volume limit spontaneous SUSY breaking is impossible provided
$\lambda^2/m^2\ll 1$. Indeed, one can show that $\lambda^2/m^2$ is
the genuine dimensionless expansion parameter in the theory at hand.
At small $\lambda^2/m^2$ the theory is weakly  coupled. Since there
are no massless fields in the Lagrangian, (and they cannot appear as
bound states in the weak coupling regime), there is no appropriate
candidate to play the role of Goldstino, and,  hence, supersymmetry
must be realized linearly.

The Witten index for the sine-Gordon model also vanishes. This can be
checked by putting the theory in a finite box. The resulting quantum
mechanics has two  supersymmetric vacua -- one bosonic and one
fermionic -- provided the wave functions considered are periodic. One
can lift these two vacua from zero by imposing Bloch boundary
conditions on the wave functions, with the angle $\theta\neq 0$.

\section{Ultraviolet and infrared regularization}
\label{regularization}
\setcounter{equation}{0}

\subsection{Superalgebra (unregularized)}
\label{ssec:algebra}

We start with the operator construction which corresponds to  the
classical level of consideration. The SUSY algebra is built on the
supercharges $Q_\alpha$, defined as
\beq
Q_\alpha=\int \! {\rm d}z\, J^{\,0}_\alpha\, ,
\eeq
where the conserved  supercurrent is
\beq
J^\mu = (\partial_\nu \phi )\gamma^\nu \gamma^\mu\psi -
i F\gamma^\mu\psi\, .
\label{cosc}
\eeq
Using the canonical commutation relations~\footnote{Here and below
$\dot
\phi$
should be understood as the canonical momentum $\pi_\phi$
conjugate
to $\phi$.}
\begin{equation}
\left[ \phi (t,\, z), \dot \phi (t,\, z^\prime) \right]= i\delta
(z-z^\prime)\,,\quad \left\{ \psi_\alpha (t,\, z),  \bar \psi_\beta
(t,\,
z^\prime)\right\}= \left(\gamma^0\right)_{\alpha \beta}\delta
(z-z^\prime)
\label{canon}
\end{equation}
one finds the anticommutator,
\begin{equation}
\left\{J_{\alpha}^\mu, \bar Q_\beta \right\}
=2\,(\!\gamma_\nu)_{\alpha\beta} \,\vartheta^{\mu\nu} +
2i\,(\!\gamma^5)_{\alpha\beta}\, \zeta^\mu
\;,
\label{celoc}
\end{equation}
where $\vartheta^{\mu\nu}$  is the energy-momentum tensor
\begin{equation}
\vartheta^{\mu\nu}= \partial^\mu \!\phi \,\partial^\nu \!\phi +\frac
12
\bar\psi \gamma^\mu \, i \partial^\nu \!\psi - \frac 12\, g^{\mu\nu}
\left[ \partial_\gamma \phi \, \partial^\gamma \!\phi -F^2\right]
\label{enmomten}
\;,
\end{equation}
and $\zeta^\mu$ is the conserved topological current,
\begin{equation}
\zeta^\mu=\epsilon^{\mu\nu} \partial_\nu {\cal W}~.
\label{topcurr}
\end{equation}
Symmetrization (antisymmetrization) over the bosonic (fermionic)
operators in the products is implied in the above expressions. Also,
the notation,
$
\gamma^5 = \gamma^0\gamma^1= -\sigma_1
$ and $F=-{\cal W}^\prime$, is used.

Integrating the $\mu=0$ component of \mbox{Eq.\ (\ref{celoc})}  over
space gives
the SUSY algebra:
\begin{equation}
\{{Q}_\alpha, \bar Q_\beta\}
=2\,(\!\gamma^\mu)_{\alpha\beta} \,P_\mu +
2i\,(\!\gamma^5)_{\alpha\beta}\,
 {\cal Z}\;.
\label{ce}
\end{equation}
Here $P_\mu=\int\! {\rm d} z
\,\vartheta_{0\mu}$ are operators of the total energy and
momentum,
and ${\cal Z}$ is the central charge,
\begin{equation}
{\cal Z} = \int \! {\rm d}z \,\zeta^0= \int \! {\rm d}z\, \partial_z
{\cal W }(\phi) =
{\cal  W}[\phi (z=\infty )]-
 {\cal W}[\phi (z=-\infty )]\, ,
\label{central}
\end{equation}
which coincides with the topological one. In the theories (\ref{minlag})
with several degenerate vacua  the central charge ${\cal Z} $ is
nontrivial for the soliton configurations. Thus, the ${\cal N}=1$
superalgebra gets centrally extended~\cite{WO}.

Let us emphasize that \mbox{Eq.\ (\ref{celoc})} establishes
supersymmetry relations between the supercurrent $J_\mu$, the
energy-momentum tensor $\vartheta_{\mu\nu}$, and the topological
current  $\zeta_\mu$, and for this  reason cannot be changed by
quantum corrections. The explicit expressions~(\ref{cosc}),
(\ref{enmomten}) and  (\ref{topcurr}) for $\vartheta_{\mu\nu}$ and
$\zeta_\mu$ are classical ones, quantum corrections will change them.

\subsection{Anomaly in the central charge}
\label{ssec:anomaly}

The structure of the centrally extended superalgebra (\ref{ce}) is quite
general. At the same time, the particular expression for the central
charge depends on the dynamics of the model under consideration.
Equation (\ref{central}) is obtained at the classical level. It is easy
to see that quantum corrections do modify it. Here we will present a
simple argument demonstrating the emergence of an anomalous term in the
central charge. We also discuss its physical meaning. A detailed direct
calculation will be  carried out after we introduce the ultraviolet
regularization by higher derivatives.

To begin with, let us consider $\gamma^\mu J_\mu$ where
$J_\mu$ is the supercurrent defined in \mbox{Eq.\ (\ref{cosc})}. This
quantity is related to the superconformal properties of the model
under consideration.  At the classical level
\beq
\left( \gamma^\mu J_\mu\right)_{\rm class} =-2i\, F\psi
\,.
\label{gsc}
\eeq
Note that the first term in the supercurrent~(\ref{cosc}) gives no
contribution in Eq. (\ref{gsc}) due to the fact that in two dimensions
$\gamma_\mu\gamma^\nu\gamma^\mu = 0$.

 Multiplying \mbox{Eq.\ (\ref{celoc})} by $\gamma_\mu$ from the
left we get the supertransformation of $\gamma_\mu J^\mu$,
\beq
\frac{1}{2} \, \left\{ \gamma^\mu J_\mu\,, \bar Q
\right\} = \vartheta^\mu_\mu +i \gamma_\mu \gamma^5
\,\zeta^\mu
\, .
\label{clgscst}
\eeq
This equation  establishes the supersymmetric relation between
$\gamma^\mu J_\mu$, $\vartheta^\mu_\mu $ and $\zeta^\mu$ and,
as was mentioned above,  remains valid with quantum corrections
included. But expressions for these  operators could be changed.
Classically  the trace of the energy-momentum tensor is
\begin{equation}
\left(\vartheta^\mu_{\,\mu}\right)_{\rm class}=F^2 + \frac 12\,{\cal
W}^{\prime\prime}\,\bar\psi\psi \;,
\end{equation}
as follows from \mbox{Eq.\ (\ref{enmomten})}. The zero component
of $\zeta^\mu$  in the second term  classically coincides with the
density of the central charge, $\partial_z{\cal W}$, see \mbox{Eq.\
(\ref{topcurr})}. It is seen that the trace of the energy-momentum
tensor and the density of the central charge appear in this relation
together.

It is  well-known that in renormalizable theories with
ultraviolet logarithmic divergences, both the trace of the
energy-momentum tensor and $\gamma^\mu J_\mu $ have
anomalies. We will use this fact, in conjunction  with \mbox{Eq.\
(\ref{clgscst})}, to establish the general form of the anomaly in the
density of the central charge.

To get some idea of the anomaly, it is convenient to use the dimensional
regularization. If we assume that the number of dimensions $D= 2 -
\varepsilon$ rather than $D=2$ the first term in \mbox{Eq.\
(\ref{cosc})} does generate a nonvanishing contribution to $\gamma^\mu
J_\mu$, proportional to $(D-2) (\partial_\nu\phi )\,\gamma^\nu\psi $. At
the quantum  level  this operator  gets an ultraviolet logarithm (i.e.
$(D-2)^{-1}$ in dimensional regularization), so that $D-2$ cancels, and
we are left with an anomalous term in $\gamma^\mu J_\mu$.

To do the one-loop calculation we apply here (and in many instances
below) the background field technique: we substitute the field $\phi$
by its background and quantum parts, $\phi$ and $\chi$, respectively,
\begin{equation}
\phi \longrightarrow \phi+\chi \,.
\label{backsubst}
\end{equation}

Specifically, for the anomalous  term in $\gamma^\mu J_\mu$,
\begin{equation}
\left( \gamma^\mu J_\mu\right)_{\rm anom} \!=
(D-2)\,  (\partial_\nu\phi )\, \gamma^\nu\psi=
-(D-2)\,  \chi  \gamma^\nu\partial_\nu\psi=
i\,(D-2)\, \chi \, {\cal W}''(\phi+\chi)\,\Psi
\,,
\end{equation}
where an integration by parts has been carried out, and a total
derivative term  is omitted (on dimensional grounds it vanishes in
the limit $D=2$). We also used  the equation of motion for  the $\psi$
field.
The quantum field $\chi$ then forms a loop and we get for the
anomaly,
\begin{eqnarray}
\left( \gamma^\mu J_\mu\right)_{\rm anom}\!\!\!&=&\!\!
i\,(D-2)\,\langle 0|\chi^2|0\rangle\, {\cal W}'''(\phi)\,\psi
=-(D-2)\! \int \!\frac{{\rm d}^D p}{(2\pi)^D} \, \frac{1}{p^2-m^2} \;
{\cal W}'''(\phi)\,\psi \nonumber \\
&=&\!\frac{i}{2\pi}{\cal W}'''(\phi)\,\psi
\,.
\end{eqnarray}

The supertransformation of the anomalous term in
$\gamma^\mu J_\mu$ is
\beq
\frac{1}{2} \, \left\{ \left(\gamma^\mu
J_\mu \right)_{\rm anom},\bar Q
\right\} \,  =\, \left(\frac{1}{8\pi}\, {\cal W}''''\bar\psi\psi
-\frac{1}{4\pi}{\cal W}'''F\right)
+i \gamma_\mu \gamma^5 \epsilon^{\mu\nu}\partial_\nu
\left(\frac{1}{4\pi}{\cal W}''\right)\, .
\eeq
The first term on the right-hand side is the anomaly in the trace of
the energy-momentum tensor, the second term
represents the anomaly in the topological current, the corrected
current has the form
\beq
\zeta^\mu = \epsilon^{\,\mu\nu}\partial_\nu \left( {\cal W}+
\frac{1}{4\pi}\, {\cal
W}''\right)\, .
\label{ccdanoc}
\eeq
Consequently, the operator of the central charge
\beq
{\cal Z} =\left( {\cal W}+ \frac{1}{4\pi}\, {\cal
W}''\right)_{z=+\infty} -\left( {\cal W}+ \frac{1}{4\pi}\, {\cal
W}''\right)_{z=-\infty}
\, ,
\label{anocc}
\eeq
is the difference of ${\cal W}+ ({\cal W}''/4\pi)$
 at $z=\pm \infty$.
Other anomalies in the supermultiplet  together with their classical
counterparts  have the following form,
\begin{eqnarray}
\gamma^\mu J_\mu &=&2i\,\left({\cal W}' +\frac{{\cal
W}'''}{4\pi}\right)\psi\,, \nonumber \\
\vartheta^\mu_{\,\mu}&=&-F\left({\cal W}' +\frac{{\cal
W}'''}{4\pi}\right) + \frac 12\,\left({\cal W}'' +\frac{{\cal
W}''''}{4\pi}\right)\,\bar\psi\psi \;.
\label{counterparts}
\end{eqnarray}

 The physical meaning of the anomalous term in the central charge
becomes quite apparent from the consideration above.
The supersymmetry relates the anomaly in the topological current
with a response of the system to the scale transformations,
\beq
x\to (1-\rho ) x\, , \,\,\,  \theta
\to (1-(\rho /2))\theta\, , \,\,\,  \Phi \to \Phi \, .
\label{sctr}
\eeq
The kinetic term of the action (\ref{minac})
is invariant. The non-invariance of the classical superpotential term
is associated with ${\cal W}$. This is not the end of the story,
however.
To reveal the anomaly it is convenient, simultaneously with
(\ref{sctr}),  to rescale the parameters in the superpotential
in accordance with their canonical dimensions. For instance, in the SPM
model $m\to (1+\rho )m\, , \,\,\, \lambda
\to (1+\rho )\lambda$, and in the SSG model $m\to (1+\rho )m\, ,
\,\,\,
v\to v$.
Then,
\beq
{\cal W} \to (1+\rho ){\cal W}\,,
\label{trlaw}
\eeq
and the classical action is invariant (the canonical
dimension of ${\cal W}$ in two dimensions is unity).
The quantum action is {\em not} invariant, due to the occurrence
of a new dimensional parameter, the ultraviolet regulator mass,
see e.g. Eq. (\ref{deltaw}). As it follows from Eq. (\ref{deltaw}),
the non-invariant contribution is ${\cal W}''/4\pi$. This term gives
rise to the anomaly in the central charge density.

It is instructive to compare the above pattern with the anomaly in
the central charge in the four-dimensional Wess-Zumino models,
discussed previously in the literature \cite{CS}.
In four dimensions, the central charge is proportional to the jump in
\beq
3{\cal W} - \sum_i \Phi_i \frac{\partial{\cal W}}{\partial\Phi_i}
+\frac{1}{8}\sum_i \gamma_i\bar{D}^2(\bar\Phi_i \Phi_i )\, ,
\label{cc4d}
\eeq
where the summation runs over all chiral superfields $\Phi_i$ on
which the given Wess-Zumino model is built. Here $\gamma_i$ is the
anomalous dimension of the field $\Phi_i$. The first two terms are
purely classical. The coefficient 3 in front of ${\cal W}$ is due
to the fact that the canonical dimension of the superpotential in the
four-dimensional Wess-Zumino model is three. The second term
reflects the canonical dimension of the superfield $\Phi$ which is
equal to unity in four dimensions. (This is in contrast to two
dimensions, where the canonical dimension of $\Phi$ vanishes.)
The  first two terms in Eq. (\ref{cc4d}), combined together,
vanish for purely cubic superpotentials. The reason is quite clear --
if the superpotential in the four-dimensional Wess-Zumino model is
a cubic polynomial, such a model is classically scale invariant, so the
classical part in the central charge must also vanish.

The last term in Eq. (\ref{cc4d}) is the quantum anomaly.
It reflects the fact that the field $\Phi$ in loops acquires an
anomalous dimension, through a logarithmic renormalization of the
kinetic term. In fact, due to the non-renormalization theorem
\cite{GRS}, the $Z$ factor for the kinetic term is the only
renormalization occurring in the theory. Hence, the term
proportional to the anomalous dimensions $\gamma_i$ is the only
anomalous term. In two dimensions, the kinetic term receives no
logarithmically divergent renormalizations, and the anomalous
dimension of $\Phi$ vanishes. However, unlike four dimensions, the
superpotential itself gets renormalized. This is the origin of
the ${\cal W}''/4\pi$ anomalous term  in Eq. (\ref{ccdanoc}) in two
dimensions. In four dimensions the anomalous term is a full
superderivative, its  expectation value over the supersymmetric
vacuum vanishes  automatically. Therefore, it plays no role in the
computation of the  central charge for the domain walls \cite{CS}. For
the very  same reason, there is no anomalous correction in the central
charge in the two-dimensional soliton problems with ${\cal N}=2$. At
the same time, the expectation value of the ${\cal W}''$ anomalous
term in two dimensions by no means vanishes. The contribution of the
anomaly to the central charge (and, hence, to the mass of the BPS
saturated solitons) is absolutely essential for the overall
self-consistency of the analysis.

\subsection{Representations of the centrally extended algebra}
\label{ssec:represcea}

Let us consider the representation of the algebra~(\ref{ce}) in the
soliton sector.  In the rest frame, where  $P_z=0$, we have
\begin{equation}
Q_1^2=H+{\cal Z}\;,\quad Q_2^2=H-{\cal Z}\;.
\label{represent}
\end{equation}
The supermultiplet consists of two degenerate states: the bosonic
$|\,{\rm sol}_{\,b}\rangle$ and fermionic  $ |\,{\rm sol}_{\,f}\rangle$
kinks.\footnote{Actually, the supermultiplet is reducible: the states $(|{\rm sol}_b \rangle \pm |{\rm sol}_f \rangle)/\sqrt{2}$ present two one-dimensional representations of superalgebra.} The supercharge $Q_1$ connects them,
\begin{equation}
Q_1 |\,{\rm sol}_{\,b}\rangle =\sqrt{2 M} |\,{\rm
sol}_{\,f}\rangle\;,\quad
Q_1 |\,{\rm sol}_{\,f}\rangle =\sqrt{2 M} |\,{\rm sol}_{\,b}\rangle\,.
\label{qone}
\end{equation}
The second supercharge $Q_2$ annihilates both soliton states,
\begin{equation}
Q_2\,|\,{\rm sol}\rangle =0\;.
\label{singlet}
\end{equation}
This equation is the operator form of the BPS saturation condition.
Previously we discussed both features, \mbox{Eq.\ (\ref{qone})} and
\mbox{Eq.\ (\ref{singlet})}, at the level of the classical solution.

Note, that an analogue of the  condition~(\ref{singlet}) in four
dimensions is related to multiplet shortening.  This multiplet
shortening is possible only if there is a supercharge, which similarly
to $Q_2$, annihilates some states, as in \mbox{Eq.\ (\ref{singlet})},
and its square is related to $H-{\cal Z}$, as in \mbox{Eq.\
(\ref{represent})}.  In fact, the multiplet shortening is nothing but a
signature (sufficient condition) of the existence of an invariant
subalgebra.  Although we do  not have multiplet shortening in two
dimensions, the invariant subalgebra formed by $Q_2$ and $Q_2^2$ still
exists. The relation~(\ref{singlet}) is enough to guarantee the absence
of quantum corrections  to $\langle{\rm sol}|H-{\cal Z}|\,{\rm
sol}\rangle$.

Can the soliton become non-saturated after the quantum corrections are
switched on? The answer is no. In any supersymmetric theory,  if $Q_2$
annihilates a state at the tree level, it will continue to do so to any
finite order if the coupling constant is small (for a review see
e.g.~\cite{EW}). The proof is based on the fact that the  number of
degenerate states in the multiplet cannot be increased from two to four
in the weak coupling regime. Indeed, if both supercharges, $Q_1$ and
$Q_2$, acted nontrivially one could construct two fermion states,
$Q_1|\,{\rm sol}_{\,b}\rangle$ and $Q_2|\,{\rm sol}_{\,b}\rangle$. Then
the second bosonic state would be obtained as $Q_1Q_2|\,{\rm
sol}_{\,b}\rangle$. Since at the classical level there is no extra pair
of degenerate states it cannot appear in  perturbation theory. Thus, the
soliton is protected from losing its saturated nature.

The condition of BPS saturation~(\ref{singlet}) implies that
\begin{equation}
\langle {\rm sol}|H-{\cal Z}|\,{\rm sol}\rangle=0
\label{nocorr}
\end{equation}
is valid to any order in  perturbation theory.
In a sense, $H-{\cal Z}$ can be viewed   as a Hamiltonian for  the
problem  with the boundary effects  taken into consideration. Then
\mbox{Eq.\ (\ref{nocorr})} expresses the complete cancellation of the
quantum corrections between bosons and fermions. It relates the
soliton mass  to the central charge,
\begin{equation}
M\equiv\langle {\rm sol}|H|\,{\rm sol}\rangle=\langle {\rm
sol}|{\cal Z}|\,{\rm sol}\rangle
\;.
\label{masscentr}
\end{equation}

\subsection{Ultraviolet regularization by higher derivatives}
\label{ssec:uvreg}
We use higher derivatives to regularize the theory in the ultraviolet.
The main advantage of the method is the explicit preservation of
supersymmetry at every step. To simplify the construction of the
canonical formalism we introduce only spatial derivatives for the
regularization, the Lorentz noncovariance of the procedure does not
lead to a problem.

In regularized form  the action~(\ref{minac})
becomes
\begin{equation}
S =i\int {\rm d}^2\theta\, {\rm d}^2x \left\{ \frac{1}{4}\bar D_\alpha
\Phi \left( 1-\frac{\partial_z^2}{M_r^2}\right) D_\alpha\Phi + {\cal
W}(\Phi
)\right\}
\,,
\label{minacreg}
\end{equation}
where $M_r$ is the regulator mass.
In the component form we arrive at the following Lagrangian
\begin{eqnarray}
{\cal L}&=& \frac{1}{2}\left\{ \partial_\mu\phi \left(
1-\frac{\partial_z^2}{M_r^2}\right) \partial^\mu\phi +\bar\psi  \left(
1-\frac{\partial_z^2}{M_r^2}\right) i\!\not\!\partial \psi +F \left(
1-\frac{\partial_z^2}{M_r^2}\right)\! F  \right. \nonumber\\[0.2cm]
&&\left.+ 2\,{\cal W}'(\phi)F -{\cal
W}''(\phi)\bar\psi\psi \right\}\, .
\label{minlagreg}
\end{eqnarray}
The auxiliary field $F$ can be eliminated from the Lagrangian by
virtue of the following formula:
\begin{equation}
 F= - \left( 1-\frac{\partial_z^2}{M_r^2}\right)^{-1} \! {\cal
W}^{\prime}
\,.
\label{fw}
\end{equation}

In  perturbation theory near the flat vacuum the bosonic and
fermionic propagators are
\begin{eqnarray}
\frac{M_r^2}{M_r^2+p_z^2} \,\, \frac{1}{p^2 -({\cal
W}_0^{\prime\prime})^2\,\left[1+(p_z^2/M_r^2)\right]^{-2}}\,,&&
\mbox{~~(bosonic)}
\nonumber\\
\frac{M_r^2}{M_r^2+p_z^2} \, \,\frac{1}{\not \! p -{\cal
W}_0^{\prime\prime}\,\left[1+(p_z^2/M_r^2)\right]^{-1}}\,,&&
\mbox{~~(fermionic)}
\label{propagators}
\end{eqnarray}
where ${\cal W}_0^{\prime\prime}$ is the second derivative
evaluated at the vacuum value of the field $\phi$.  The regularized
propagators  acquire an extra factor $M_r^2/(M_r^2+p_z^2)$ making
the integrals over $p$ convergent.  Strictly speaking, for the
quadratically divergent integrals a higher power of  the same factor is
needed -- one can achieve this simply by increasing the  power of $(1
-\partial_z^2/M_r^2)$ in \mbox{Eq.\ (\ref{minacreg})}.

Proceeding with canonical quantization in the usual way we get
the canonical commutation relations:
\begin{eqnarray}
\left[ \phi (t,\, z), \left(1 - \frac{\partial_{z'}^2}{M_r^2}\right)\dot
\phi (t,\,
z^\prime)
\right]= i\delta\,
(\! z-z^\prime)\,,\nonumber\\
\left\{ \psi_\alpha (t,\, z),  \left(1 -
\frac{\partial_{z'}^2}{M_r^2}\right)\bar \psi_\beta (t,\,
z^\prime)\right\}= \left(\gamma^0\right)_{\alpha \beta}\delta\,
(\! z-z^\prime)\,,
\label{canonreg}
\end{eqnarray}
and the
regularized Hamiltonian density ${\cal H}$,
\begin{eqnarray}
{\cal H}&=&\dot \phi \left(1 -
\frac{\partial_{z}^2}{M_r^2}\right)\dot \phi
+ \frac 12 \left[  \left(1 -
\frac{\partial_{z}^2}{M_r^2}\right) \bar \psi\right] \dot \psi -{\cal
L}
\label{hdensity} \\[0.1cm]
&=&\frac{1}{2}\left\{ \dot\phi \left(
1-\frac{\partial_z^2}{M_r^2}\right) \dot\phi + \partial_z\phi \left(
1-\frac{\partial_z^2}{M_r^2}\right) \partial_z\phi  - {\cal W}' F
-i\bar\psi\left(1-\frac{\partial_z^2}{M_r^2}\right)\gamma^1\psi
+{\cal W}'' \bar\psi\psi \right\} \,.\nonumber
\end{eqnarray}

It is also simple to construct the regularized conserved supercurrent,
\begin{equation}
J^\mu=\left[\left(1 - \frac{\partial_{z}^2}{M_r^2}\right)\left(\not\!
\partial \phi
+iF\right)\right] \gamma^\mu \psi +
\frac{1}{M_r^2}\,\delta^\mu_1\left\{
-\partial_z (\!\not\!\partial\phi \not\!\partial\psi) +iF\, \partial_z\!
\not\!\partial\psi -i \,(\partial_z  F)\!\not\!\partial\psi\right\}\;.
\label{superregcur}
\end{equation}
Using the canonical commutation relations presented above one can
calculate the commutator of this regularized current with the
regularized supercharge. This is a relatively straightforward although
tedious exercise. In this way we  get regularized expressions for the
energy-momentum tensor $\vartheta_{\mu\nu}$ and the topological current
$\zeta_\mu$. We will analyze them to get the anomaly in the next
subsection.

\subsection{Calculation of the anomaly by higher-derivative
regularization}
\label{ssec:cahdr}

The particular anticommutator which determines the regularized
topological
current $\zeta^\mu$ is
\begin{equation}
\zeta^\mu= -\frac{i}{4} \,{\rm Tr}\left( \left\{ J^\mu, \,\bar Q
\right\}
\gamma^5\right)\, ,
\label{keyco}
\end{equation}
where $J^\mu$ is presented in \mbox{Eq.\ (\ref{superregcur})}.
The complete expression is rather cumbersome, therefore we will
limit ourselves
to  the time
component $\zeta^{\,0}$ which is sufficient for extracting the
anomaly,
\begin{equation}
\zeta^{\,0}= \partial_z\left\{ {\cal W} +\frac{1}{2M_r^2}\left[
(\partial_z\phi )
\,\partial_z F
-(\partial_z^2 \phi )\, F \right]\right\}
\,,
\label{topreg1}
\end{equation}
where $F$ is given in \mbox{Eq.\ (\ref{fw})}.
The term $\partial_z {\cal W} $  on the right-hand side is the
classical topological
charge density.  The additional terms are due to the regularization,
formally they
vanish as $1/M_r^2$ in the limit of the large regulator mass.
However, the loop
integration yields $M_r^2$ in the numerator, so a finite additional
term survives.

Using the background field decomposition~(\ref{backsubst}),
$\phi\longrightarrow \phi_0 +\chi$,  we
expand the right-hand side of \mbox{Eq.\ (\ref{topreg1})} up
to the second order in the quantum field $\chi$  and  keep only
those terms which survive in the limit $M_r\to \infty$,
\begin{equation}
\frac{1}{2M_r^2}\left[ (\partial_z\phi )\,
\partial_z F
-(\partial_z^2 \phi ) \,F\right]
=\,\frac{1}{M_r^2}\,{\cal W}^{\prime\prime} \left \langle \partial_z
\chi \left(1
-
\frac{\partial_{z}^2}{M_r^2}\right)^{-1} \!\partial_z \chi \right
\rangle
\,.
\label{topreg2}
\end{equation}
Strictly speaking, the Green's function of the quantum field $\chi$
depends on the background $\phi_0$; see Eq. (\ref{propagators}). This
dependence drops out, however, and the result for the $\chi$ loop does
not depend on the choice of the background field $\phi_{\,0}$ since the
loop integral is saturated in  the domain of  virtual momenta $p_z \sim
M_r$. Other domains are irrelevant in the limit $ M_r\to\infty$. Thus we
can use the bosonic propagator from \mbox{Eq.\ (\ref{propagators})},
neglecting the  $({\cal W}_0^{\prime\prime})^2$ term in its denominator,
\begin{equation}
\left \langle \partial_z \chi \left(1 -
\frac{\partial_{z}^2}{M_r^2}\right)^{-1} \partial_z \chi \right \rangle
=
i\int\! \frac{{\rm d}^2 p}{(2\pi)^2}\,
 p_z^2 \,\frac{1}{p^2
\left[1+(p_z^2/M_r^2)\right]^2}=\frac{M_r^2}{4\pi}
\,.
\label{topreg3}
\end{equation}
As a result, combining \mbox{Eqs.\ (\ref{topreg1}), (\ref{topreg2}),
(\ref{topreg3})},
we arrive at
\begin{equation}
\zeta^{\,0}= \partial_z\left\{ {\cal W} +\frac{{\cal
W}^{\prime\prime}}{4\pi}
\right\}\,,
\end{equation}
which exactly coincides with \mbox{Eq.\ (\ref{ccdanoc})} in the time
component. Dimensional regularization and that by higher
derivatives lead to one and the same expression for the anomaly
in the central charge.

Instead of computing the commutator (\ref{keyco}) we can suggest
an alternative procedure leading directly to  Eq. (\ref{topreg1}). To
this end  it is enough to note that the
regularized Hamiltonian density given by \mbox{Eq.\ (\ref{hdensity})}
can be identically rewritten as follows
\begin{eqnarray}
{\cal H}&=&\zeta^{\,0} + \frac{1}{2}\left\{ \dot\phi \!\left(
1-\frac{\partial_z^2}{M_r^2}\right)\! \dot\phi + (\partial_z\phi+F)\!
\left(
1-\frac{\partial_z^2}{M_r^2}\right) (\partial_z\phi+F)\right.
\nonumber\\[0.2cm]
&&\left.
-i\bar\psi\left(1-\frac{\partial_z^2}{M_r^2}\right)\gamma^1\psi
 +{\cal W}'' \bar\psi\psi \right\}\,,
\label{energy0}
\end{eqnarray}
where the topological charge density $\zeta^{\,0}$ is the same as in
\mbox{Eq.\ (\ref{topreg1})} and $F$ is defined by
\mbox{Eq.\ (\ref{fw})}.

In the subsequent sections, the very same anomaly will be confirmed,
additionally, by two alternative analyses: of the effective
one-loop superpotential; and models with an extra superfield.
The extra superfield  introduced in the framework of softly broken
extended supersymmetry serves as a Lorentz-invariant regulator
when its mass is taken to be large.

\subsection{One-loop nature of the anomaly}
\label{ssec:highloops}

The anomalous term ${\cal W}''/4\pi$ was obtained at one loop. A natural
question is what happens at higher  loop orders. In this short
subsection  we will argue that  the expression~(\ref{ccdanoc}) is {\em
exact} -- no higher loop corrections emerge. This should be understood 
as an operator relation.

The anomaly comes from the ultraviolet range, as we have demonstrated by
explicit calculations above. In this range one can consider the
superpotential term in the action~(\ref{minac}) as a perturbation to the
first (kinetic) term, which describes massless free fields. The
superpotential ${\cal W}$ is proportional to the mass $m$,  and $m/M_r$
is our small parameter.  (In the polynomial model the coupling constant
$\lambda \sim m v$.)

Consider, for instance, the operator $\gamma^\mu J_\mu$. At zeroth order
in ${\cal W}$ this operator vanishes, in the first order in ${\cal W}$
it is proportional to ${\cal W}$ (more precisely,  ${\cal W}$ and its
derivatives).  All higher loops, i.e. higher orders in ${\cal W}$, will
result in an extra power of $m/M_r$ as far as contributions from
ultraviolet domain are concerned. Thus, the first loop is the only one
which produces the anomaly.

Does this mean that if one calculates the Feynman graphs for $\gamma^\mu
J_\mu$ the only graph that contributes is one loop? The answer is no.
Certainly, multi-loop contributions are present too, but they all
represent  the  {\em matrix element} of the operator on the right-hand
side of Eq. (\ref{counterparts}). The fact that all nonvanishing
multi-loop corrections are of infrared origin and, thus, must be referred
to the matrix element of the appropriate operator, is obvious on
dimensional grounds. Thus, the situation with the anomaly in
two-dimensional models under discussion is a perfect parallel to that in
SUSY gluodynamics in four dimensions where the three ``geometric"
anomalies, considered as operator relations,  are one-loop exact; all
higher-order corrections come as matrix elements of the operator Tr$W^2$
in the given background field \cite{MSAV}.

Confirmations of the one-loop nature of the anomaly are provided by
calculations in the SSG model presented in Sec. \ref{2loops}.

\subsection{Infrared regularization and fermion-boson cancellations in
local quantities}
\label{ssec:local}

To discretize the spectrum of modes let us put the system in a large
box, i.e. impose boundary conditions at $z=\pm L/2$.
It is convenient to choose them in a form which is compatible
with the residual supersymmetry~(\ref{halfsusytr}). The boundary
conditions
we are imposing are
\begin{eqnarray}
&&\left. \left(\partial_z\phi +F \right) \right|_{z=\pm L/2}=0\,,
\nonumber\\[0.2cm]
&&\left. \left(\partial_z +F^\prime\right)
\psi_2\right|_{z=\pm  L/2}=0 \,,
\nonumber\\[0.2cm]
 && \left. \psi_1\right|_{z=\pm  L/2}=0\,.
\label{boundary}
\end{eqnarray}
It is easy to verify the invariance of these conditions under
transformations~(\ref{halfsusytr}). They are also consistent with the
classical solutions, both for the flat vacuum and for the kink. In
particular,
the soliton solution $\phi_0$  satisfies to
$\partial_z\phi +F =0$ everywhere, and boundary
conditions~(\ref{boundary}) do not  deform it; we can place the kink in
the box.

In Sec. \ref{ssec:represcea} the exact equality between the soliton mass
and the expectation value of the central charge was discussed. Here we
will consider a stronger statement, namely the exact equality between
the energy density inside the soliton and the expectation value  of the
central charge density. This equality is valid to any order in
perturbation theory and holds {\em locally}, for any $z$.

Let us use the expression~(\ref{energy0}) for the
energy density $\cal H$ in the following form,
\begin{equation}
{\cal H} = \zeta^{\,0}+ \frac{1}{2}\left\{(\dot \phi)^2
+(\partial_{\,z} \phi+F)^2 +i\psi_2(\partial_{\,z} - F^\prime)\psi_1
+ i \psi_1(\partial_{\,z} +F^\prime)\psi_2\right\}\,,
\label{energy1}
\end{equation}
where $\psi_{1,2}$ are the components of the fermion field
$\psi_\alpha$. We have omitted for simplicity the $1/M_r^2$ terms on the
right-hand side as they do not change the consideration.
Equation~(\ref{energy1}) is the exact operator form. In the classical
limit, where $\psi=0$, the expression is positive-definite. Its minimum
is achieved provided the  field $\phi$ is chosen to be time-independent
and to satisfy the BPS condition $\partial_{\,z} \phi+F=0$. Then the
soliton mass is given by the integral over the first term in \mbox{Eq.\
(\ref{energy1})}, i.e. by the central charge $\cal Z$. But now we want
to analyze the relation (\ref{energy1}) in its local form.

 For the classical kink solution  the difference ${\cal H}-\zeta^{\,0}$
obviously vanishes at any $z$. This vanishing persists at the quantum
level,
\begin{equation}
\langle {\rm sol}\,| {\cal H}(x) - \zeta^{\,0} (x)|\,{\rm
sol}\rangle=0\,.
\label{locenergy}
\end{equation}
Indeed, the operator ${\cal H}(x) - \zeta^{\,0} (x)$ can be
presented as the anticommutator \cite{HY}
\begin{equation}
\{Q_2,J^0_2(x)\}= {\cal H}(x) -\zeta^{\,0} (x)\,,
\end{equation}
which vanishes upon averaging, since $Q_2|\,{\rm sol}\,\rangle=0$. As a
result, not only do the quantum corrections to the kink mass coincide
with  those for the central charge, but  the energy density distribution
coincides with the central charge density to all orders.

It is instructive to check in more detail how the cancellation works
in  the one-loop approximation. To this end we expand the bosonic field
$\phi$ near the classical soliton background $\phi_{\,0}$,
\begin{equation}
\phi=\phi_{\,0}+\chi\,.
\end{equation}
In the quadratic in the  quantum fields $\chi$ approximation th difference
${\cal H}-\zeta^{\,0} $ takes the following form:
\begin{equation}
\left[{\cal H}-\zeta^{\,0}\right]_{\rm quad} =
\frac{1}{2}\left\{\dot \chi^2
+[(\partial_{\,z} +F^\prime)\chi]^2
+i\psi_2(\partial_{\,z} - F^\prime)\psi_1 +
i \psi_1(\partial_{\,z} +F^\prime)\psi_2\right\}\,,
\label{H2}
\end{equation}
where $F^\prime$ is evaluated at $\phi=\phi_0$ and
the subscript ``quad'' denotes the quadratic approximation.
We recall that the prime denotes differentiation over $\phi$,
$$
 F'\equiv \frac{{\rm d} F}{{\rm d}\phi}=  -\frac{{\rm d}^2 {\cal
W}}{{\rm d} \phi^2} \, .
$$
 From this expression  one can see that a convenient basis for the mode
expansion of $\chi$ and $\psi_2$ is given by the eigenfunctions $\chi_n
(z)$
of the Hermitian differential operator $L_2$,
\beq
 L_2\,\chi_n (z)=\omega_n^2\, \chi_n (z)\,,\qquad L_2= P^\dagger P\,,
\label{eigen}
\eeq
where the operators $P$ and $P^\dagger$ defined as 
\beq
P = \partial_z + F'\, , \quad  P^\dagger = -\partial_z + F'
\label{shtwo}
\eeq
are Hermitian conjugates.
The boundary conditions for the modes in the box follow from \mbox{Eq.\
(\ref{boundary})},
\begin{equation}
\left(\partial_z +F^\prime\right)
\chi_n (z=\pm  L/2)=0 \,.
\label{boundary1}
\end{equation}

As for the basis of the mode  expansion for  $\psi_1$,  it is formed by
the eigenmodes of the operator $\tilde L_2$,
\beq
 \tilde L_2\,\tilde\chi_n (z)=\omega_n^2\, \tilde\chi_n (z)\,,\qquad
\tilde L_2=P P^\dagger \,,
\label{eigen2}
\eeq
with the boundary conditions
\begin{equation}
\tilde\chi_n (z=\pm  L/2)=0 \,.
\label{boundary2}
\end{equation}

With boundary conditions (\ref{boundary1}) and (\ref{boundary2}) all
eigenvalues of the operators (\ref{eigen}) and (\ref{eigen2}) are the
same,
with the exception of the zero mode $\chi_0 \propto {\rm d} \phi_0
(z)/{\rm
d} z$ in  \mbox{Eq.\ (\ref{eigen})}.  The operator (\ref{eigen2}) has no
zero mode. Moreover, the eigenfunctions $\chi_n$ and $\tilde\chi_n$ are
algebraically related,
\begin{equation}
\tilde\chi_n=\frac{1}{\omega_n} P \,\chi_n\,, \qquad
\chi_n=\frac{1}{\omega_n} P^\dagger \,\tilde\chi_n\,.
\label{relmodes}
\end{equation}

The expansion in eigenmodes has the form,
\begin{eqnarray}
&&\chi (x) =\sum_{n\neq 0} b_{n} (t) \,\chi_n (z)\,, \quad
\psi_2 (x)=\sum_{n\neq 0} \eta_n (t)\,\chi_n (z)\,, \nonumber \\
&&\psi_1 (x)=\sum_{n\neq 0} \xi_n (t)\,\tilde\chi_n (z)\,.
\label{expansion}
\end{eqnarray}
Note that the summation does not include the zero mode $\chi_0 (z)$.
This mode is not present in $\psi_1$ at all. As for the
expansions of $\chi$ and $\psi_2$, the inclusion of the zero mode
would correspond to a shift  in the collective coordinates $z_0$ and
$\eta$
(see
Sec.~\ref{ssec:kinks}). Since we  deal with the bosonic kink centered
at
$z=0$,  these shifts must  be forbidden.

 The coefficients $a_n$, $\eta_n$ and $\xi_n$ are
time-dependent operators. Their equal time commutation relations
are
determined by the canonical commutators~(\ref{canon}),
\begin{equation}
[b_m,\dot b_n]=i\delta_{mn}\,,\quad
\{\eta_m,\eta_n\}=\delta_{mn}\,,\quad
\{\xi_m,\xi_n\}=\delta_{mn}\,.
\end{equation}
Thus, the mode decomposition reduces dynamics of the system under
consideration to quantum mechanics of an infinite set of supersymmetric
harmonic oscillators (in higher orders the oscillators become
anharmonic). The ground state of the quantum soliton corresponds to
setting each oscillator in the set to the ground state.

Constructing the creation and annihilation operators in the standard way
we find the following nonvanishing expectations values of the bilinears
built from the operators $a_n$, $\eta_n$ and $\xi_n$ in  the ground
state:
\begin{equation}
\langle \dot b_n^2\rangle_{\rm sol}=\frac{\omega_n}{2}\,,\quad
\langle  b_n^2\rangle_{\rm sol}=\frac{1}{2\omega_n}\,,\quad
\langle \eta_n \xi_n\rangle_{\rm sol}=\frac{i}{2}\,.
\label{expect}
\end{equation}
The expectation values of other bilinears obviously vanish.
Combining \mbox{Eqs.\ (\ref{H2})}, (\ref{expansion}) and
(\ref{expect})
we  get
$$
\langle {\rm sol}\,| \left[{\cal H}(x) - \zeta^{\,0}\right]_{\rm
quad}|\,{\rm
sol}\rangle=
$$
\vspace{-0.4cm}
\begin{equation}
\frac{1}{2}\sum_{n \neq 0} \left\{\frac{\omega_n}{2}\chi_n^2 +
\frac{1}{2\omega_n} [(\partial_{\,z} +F^\prime)\chi_n]^2
-\frac{\omega_n}{2}\chi_n^2 - \frac{1}{2\omega_n} [(\partial_{\,z}
+F^\prime)\chi_n]^2\right\}\equiv 0\, ,
\label{H3}
\end{equation}
The four terms in the braces in \mbox{Eq.\ (\ref{H3})}  are in
one-to-one correspondence with the four terms in \mbox{Eq.\ (\ref{H2})}.
Note that in  proving the vanishing of the right-hand side we did not
perform  integrations by parts. The vanishing of the right-hand side of
(\ref{H2}) demonstrates explicitly the residual supersymmetry (i.e. the
conservation of $Q_2$) at work. The cancellation occurs in the very same
manner with a finite regulator mass $M_r$ due to the residual
supersymmetry.

Equation~(\ref{locenergy}) must be considered as a local version of BPS
saturation (i.e. conservation of a residual supersymmetry). The
nonrenormalization of the expectation value of ${\cal H} -\zeta^{\,0}$
over the soliton is a direct analog of the nonrenormalization theorem in
the instanton background leading to the exact $\beta$ functions
\cite{NSVZbeta} in supersymmetric gluodynamics. The idea of
fermion-boson cancellation of the quantum corrections in local
quantities was first formulated   in \mbox{Ref.\ \cite{HY}}. However, in
this work the problem of ultraviolet regularization was ignored.
Consequently, the quantum anomaly in the central charge density was
lost.

In general, the phenomenon of fermion-boson cancellation takes
place for the expectation value of  every local operator which can be
represented as $\{Q_2, {\cal O}(x)\}$ where ${\cal O}(x)$ is some
fermionic operator. We will use this feature later on  to calculate the
quantum corrections to the  energy distribution and to the mean
field in the kink.

\subsection{Explicit expressions for the modes and Green's functions}
\label{ssec:irreg}

In this subsection we present explicit expressions for the modes and
Green's functions in the SSG model. A nice feature is that everything is
expressible in terms of the modes and Green's functions in the flat
vacuum. A similar relation exists for the SPM model, and it is outlined
at the very end of the subsection.  Green's functions in the soliton
background are needed for the calculation of the field and energy
profiles at one loop; see Sec. \ref{sec:profile}.

Let us consider  the SSG model. In the kink background the function
$F'(z)$
which determines  the operators $P$ and $P^\dagger$ defined in Eq.\
(\ref{shtwo}) has the form
\beq
 F' = m\,\tanh\,mz\, ,
\label{shf}
\eeq
Consequently, the operators $L_2$ and $\tilde L_2$ which determine
the
mode decomposition are
\beq
L_2=P^\dagger P=
-\partial_z^2+ m^2 - {2 \, m^2 \over \cosh^2 m z}
~,
\label{shthree}
\eeq
and
\beq
{\tilde L_2} =PP^\dagger = - \partial_z^2 + m^2\, .
\label{quflat}
\eeq
The key observation is that  the
operator ${\tilde L_2}$ is a quadratic operator in the flat vacuum,
and its eigenmodes are those of free motion,
\begin{equation}
\tilde \chi_n=\sin p_n \left ( z + \frac{L}{2} \right )\,.
\end{equation}
These solutions satisfy
the boundary conditions (\ref{boundary2}) at
\beq
p_n = \frac{\pi n}{L}\, , \qquad n = 1, 2, 3, ...\, ,
\eeq
with the eigenvalues
\begin{equation}
\omega_n^2 = m^2 + p_n^2\, .
\end{equation}

As discussed in the previous subsection, see \mbox{Eq.\
(\ref{relmodes})}, all nonzero modes $\chi_n$ of  the operator $L_2$ are
algebraically expressible via modes $\tilde\chi_n$. The
two operators, $L_2$ and ${\tilde L_2}$, form a superpair
in the sense of Witten's supersymmetric quantum mechanics
\cite{EW,gk} (for a pedagogical discussion see \cite{schwabl}).
All non-zero modes of the operator ~(\ref{shthree}) are then trivially
 calculable by virtue of the algebraic relation~(\ref{relmodes}),
\beq
\chi_n (z) = \frac{1}{\omega_n}\, \left[-\partial_z + m\,\tanh\, mz
\right]
\,
\sin \! \left [ p_n  \left ( z + {L \over 2 }\right ) \right ]  \, .
\eeq
A similar relation holds for the Green's function,
\beq
G_\omega(z_1,z_2 ) \equiv i\sum_{n \neq 0} {\chi_n (z_1) \,
\chi_n (z_2) \over  \omega^2 -\omega_n^2 + i\varepsilon}~:
\eeq
\beq
G_\omega(z_1,z_2 )\!={1 \over \omega^2}
\left(\!-\partial_{z_1} + m\,\tanh\, m{z_1} \right) \!
\left(\partial_{z_2} + m\,\tanh\, m{z_2} \right)\!
\left [ {\tilde G}_{\omega}(z_1,z_2) -  {\tilde G}_{\omega=0}(z_1,z_2 )
 \right] ,
\label{gfrelat}
\eeq
where ${\tilde G}_\omega(z_1,z_2 )$ is the Green's function for the free
operator ${\tilde L}_2$. In the limit of a large box, $L\to \infty$, the
relevant expression for this Green's function is
\beq
{\tilde G}_\omega (z_1,z_2 )=-\frac{i}{2k}\exp \left (- k \, |z_1-z_2|
\right )
\label{gffree}
\eeq
with $k=\sqrt{m^2-\omega^2}$. Note that Eq.\ (\ref{gfrelat}) produces
the Green's function orthogonal to the zero mode of the operator $L_2$.
We will make further use these relations in Sec.\ \ref{sec:profile} and
\ref{sec:recurrency}.

In the general case, Witten's superpair of operators has the form
(see e.g. \cite{schwabl})
\beq
-\partial_z^2+ m^2 - {K(K-1) \, m^2 \over \cosh^2 m z} ~~~
{\rm and} ~~~
-\partial_z^2+ m^2 - {K(K+1) \, m^2 \over \cosh^2 m z}\, ,
\eeq
where $K$ is integer. We used this fact above  at $K =1$,
to relate the eigenmodes in the SSG kink background to those
of free motion. We can now repeat the trick to relate the
eigenmodes in the SPM kink background to those in the SSG kink
background, explicitly written above. Indeed, for the SPM kink,
\beq
F'(z) = m\, \tanh\,\frac{mz}{2}\, ,
\eeq
and the quadratic operator which determines the mode
decomposition for the field $\chi$ is
\beq
L_2 = P^\dagger P = -\partial_z^2 + m^2\left[
1- \frac{3}{2}\,\frac{1}{\cosh (mz/2)}
\right] \qquad (\mbox{SPM})\, .
\eeq
The sister operator is
\beq
{\tilde L}_2 =P P^\dagger  = -\partial_z^2 + m^2\left[
1- \frac{1}{2}\, \frac{1}{\cosh (mz/2)}
\right] \qquad (\mbox{SPM})\, .
\eeq
It is not difficult to see that
\beq
{\tilde L}_{2}^{\rm SPM} (z)  = \frac{1}{4}\,L_{2}^{\rm SSG} (z/2) +
\frac{3}{4}\,m^2\, .
\label{l2relat}
\eeq

\subsection{Separating soliton from the boundaries}
\label{ssec:separation}

\begin{figure}[h]
  \begin{center}
    \leavevmode
    \epsfbox{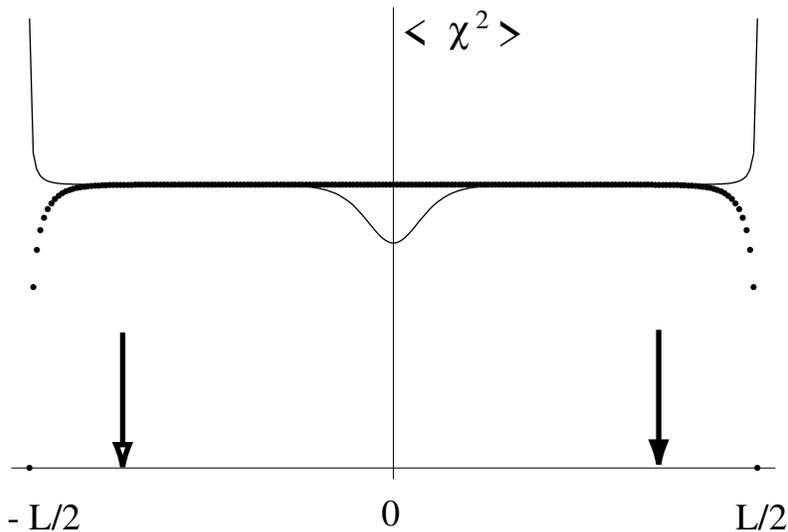}
    \caption{The quadratic average $\chi^2(z)$ in a finite box with a
finite ultraviolet regulator parameter in the SG kink background (solid
line) and for a free field (dots, coalescing into thick line away from
the edges of the box). The arrows show the boundaries of the fiducial
interval, that encloses the effects of the soliton proper as opposed to
those of the boundary conditions.}
    \label{fig:chi2}
  \end{center}
\end{figure}

We pause here to emphasize one important point concerning calculations
with boundary conditions. The relation between the eigenmodes in the two
sectors connected by the quantum-mechanical supersymmetry operator
$P$ and the exact coincidence of the spectra (with one extra zero mode in
the kink sector) only holds when the boundary conditions in the two
sectors match each other. Namely, these conditions at $z=\pm L/2$ should
be the same for
$P\chi_n/\omega_n$ in the kink sector as for ${\tilde \chi}_n$ in the
dual sector (the latter is the free motion sector in the case of the SG
model\footnote{This is {\it not} the procedure used in the standard
calculations so far: rather the same (usually periodic) boundary
conditions are imposed on $\chi_n$ and ${\tilde \chi}_n$ (see e.g. in
\cite{rajaraman}), which introduces a difference in the spectra through
a phase shift.}). Under these conditions we found the exact local in $z$
boson-fermion cancellation of the difference between the energy density
and that of the topological charge in Eq.\ (\ref{H3}). However in other
quantities, e.g. in the density of the topological charge, to be
considered in the next section, as well as in non-supersymmetric models
(like the SG model discussed in Sect.\ \ref{sec:recurrency}), the local
cancellation does not take place. In calculation of such quantities one
should clearly realize the need for separating the effects of the
soliton background from those introduced by the boundaries. The
criterion for the separation is obvious: the effects of the boundaries
are localized near the boundaries (at distances set by the mass $m$ and
by the regulator parameter $M_r$) and are carried away to infinity in
$z$ in the limit of infinitely large box: $L \to \infty$, while the
effects of the soliton proper stay at finite distances, whose scale is
set by the mass $m$. This behavior is illustrated in Fig.\
\ref{fig:chi2}, where the quadratic average of the quantum fluctuations,
$\chi^2(z)$, is shown for the SG kink background and for the flat
background, as calculated with discrete modes in a finite box with a
finite ultraviolet regulator parameter.  The spikes near the edges of
the box are the discussed artifact of the boundary conditions and are
moved to infinity in the limit of an infinite box. The soliton
background effect is the difference between the curves at the center of
the plot, and it remains fixed in the limit $L \to \infty$.

Thus the effects of the soliton proper can be separated by considering
the system in a large box up to distances well inside the box, i.e. up
to $z=\pm (L/2 -\Delta)$, where it is sufficient to choose $\Delta$ such
that $L \gg \Delta \gg m^{-1}$. This interval of $z$ where the soliton
effects already have approached their asymptotic behavior, but the
effects of the boundaries of the box are still not essential, can be
refered to as fiducial. In what follows we will imply that only the
quantities inside the fiducial interval are considered and the integral
quantities are also evaluated over the fiducial interval.

In practice the extraction of the quantities related to the averages
over the quantum fluctuations within the fiducial interval is readily
done by choosing the appropriate Green's functions for the fields of the
fluctuations. Namely, in the limit of the box and the fiducial interval
both taken to infinity these Green's functions should be chosen with the
Feynman boundary conditions, i.e. they vanish when one of the arguments
$z$ goes to either plus or minus infinity, provided that $\omega$ is shifted to the
upper complex half-plane. This is the boundary condition under which we
have specified the free-motion Green's function in Eq.\ (\ref{gffree}).
The discussed above quantum-mechanical supersymmetry then generates
the appropriate Green's functions in the soliton sector of the SG model as
well as of the SPM model (cf. Eqs.\ (\ref{gfrelat}) and (\ref{l2relat})),
which Green's functions we will be using in the next section.

\newpage
\section{The one-loop correction to the kink mass}
\label{sec:onemass}
\setcounter{equation}{0}

\subsection{Central charge and soliton mass at one loop}
\label{ssec:onecharge}

Once BPS saturation is proved the actual calculation of quantum
corrections to the kink mass becomes a simple job. Indeed, according
to \mbox{Eqs.\ (\ref{masscentr})} and (\ref{anocc})
\begin{equation}
M=\langle {\rm sol}| {\cal Z}| {\rm sol}\rangle=
\langle {\rm sol}| {\cal W}+ \frac{1}{4\pi}\, {\cal
W}''| {\rm sol}\rangle_{z=+\infty} -\langle {\rm sol}| {\cal W}+
\frac{1}{4\pi}\, {\cal W}''| {\rm sol}\rangle_{z=-\infty}\,,
\end{equation}
where the term ${\cal W}''/4\pi$ is due to the  anomaly
we discussed above in detail. At the one-loop level this term
can be substituted by its classical value at $z=\pm \infty$.
Additionally,
all we need to know is
the one-loop quantum corrections to
$\langle {\rm sol}\,|{\cal  W}|\,{\rm sol}\rangle$ at $z
\to \pm \infty$. These corrections are the same as for the flat
vacuum where the standard perturbative expansion can be used.

At $z \to + \infty$ the classical kink field $\phi_{\,0}$ goes to the
constant $\phi^+_*$. In the flat vacuum the operators $\phi$ and
${\cal  W}$ can be presented as
\begin{equation}
\phi (x)=\phi^+_* +\chi (x)\,,\quad {\cal  W}[\phi(x)]={\cal  W}_0
+\frac{1}{2}{\cal  W}^{\prime\prime}_0\,\chi^2 (x) +{\cal
O}(\chi^3)\,,
\label{flatexp}
\end{equation}
where ${\cal  W}_0={\cal  W}[\phi^+_*]$ and ${\cal
W}^{\prime\prime}_0={\cal  W}^{\prime\prime}[\phi^+_*]$ are
$c$ numbers and $\chi (x)$ is the operator. In the one-loop
approximation
\begin{equation}
\left\langle  {\cal  W} \right\rangle_0 = {\cal  W}_0
+\frac{1}{2}\,{\cal  W}^{\prime\prime}_0 \, \left\langle \chi^2 (x)
\right\rangle_{\,0}= {\cal  W}_0 +\frac{1}{2}\,{\cal
W}^{\prime\prime}_0 \,
G(x,x)\,,
\label{wcorr}
\end{equation}
where $G(x,x')$ is the propagator of the field $\chi$. Using the
regularized momentum transform  of \mbox{Eq.\ (\ref{propagators})} we get
\begin{equation}
\left\langle\,\chi^2\,\right\rangle_{\,0} = \int
\frac{{\rm d}^2 p}{(2\pi )^2}\, \frac{M_r^2}{M_r^2+p_z^2} \,\,
\frac{i}{p^2
-({\cal W}_0^{\prime\prime})^2\,\left[1+(p_z^2/M_r^2)\right]^{-2}}
=\frac{1}{4\pi}\ln
\frac{4M_r^2}{({\cal W}^{\prime\prime}_0)^2} \,,
\label{5}
\end{equation}
where $M_r$ is the ultraviolet cut off.
Equations~(\ref{wcorr}) and
(\ref{5}) give the one-loop result for
$\langle  {\cal  W} (z \to +\infty) \rangle $. The result for $\langle
{\cal
W} (z \to -\infty) \rangle $ is different only by the sign.

Thus, for the kink mass we get
\begin{equation}
M=2\,{\cal  W}_0 +\frac{{\cal  W}^{\prime\prime}_0}{4\pi}\left[
\ln \frac{4 M_r^2}{({\cal W}^{\prime\prime}_0)^2} +2\right]\,.
\label{massqone}
\end{equation}
The first term is the classical result for the mass, the one-loop
correction is given by the second term, the logarithm in the square brackets
arises from the correction to $\left\langle  {\cal  W} \right\rangle$
and the additional term 2 is from the anomaly.

It is clear that to make sense of the quantum correction we need to
formulate the renormalization procedure. i.e. to fix some observables
other than the kink mass. The mass of the light particles proves to be
appropriate. It is simple to evaluate the one-loop correction to the
fermion mass in the SSG model. It is given by the tadpole diagram and
can be written as
\begin{equation}
m_{\rm ren}= -\left\langle  {\cal  W}'' \right\rangle_0
=-{\cal  W}''_0 -\frac{1}{2}\,{\cal  W}^{\prime\prime\prime\prime}_0 \,
\left\langle \chi^2 (x)
\right\rangle_{\,0}\,,
\label{defssg}
\end{equation}
where, as above, $\left\langle \chi^2 \right\rangle_{\,0}$
appears.
At the one-loop level in the SSG model $m_{\rm ren}$ is both, the
on-shell
mass, and the renormalized parameter in the superpotential (see the next
subsection). Accounting for the fact that in the SSG model
${\cal W}=m v^2 \sin\phi/v$ and $\phi_*^+=\pi v/2$ the result ~(\ref{massqone})
for the kink mass in terms of $m_{\rm ren}$ has the form:
\begin{equation}
M_{\,{\rm SSG}}=2\,m_{\rm ren}\left[v^2 -\frac{1}{4\pi}\right]=2\,m_{\rm
ren}\,v^2 -\frac{m_{\rm ren}}{2\pi}\,.
\label{ssgmass}
\end{equation}
The logarithm term is absorb by the mass renormalization, what remains
is the anomaly term.

In the SPM model the one-loop expression for the on-shell mass is more
complicated,
it  will be obtained in Sec.\ \ref{ssec:spm}. We can introduce, however,
an off-shell renormalized mass parameter as follows:
\begin{equation}
\mu_{\rm ren}= -\left\langle  {\cal  W}'' \right\rangle_0=
2\lambda\,\phi_{\rm vac}= m+ 2\lambda \left\langle  \chi
\right\rangle_0
\,.
\label{defspm}
\end{equation}
Using the fact that $\left\langle  {\cal  W}' \right\rangle_0=0$, see
Eq.\
(\ref{doubleu}),  we can express $\left\langle  \chi \right\rangle_0$
in terms of $\left\langle  \chi^2 \right\rangle_0$ which was determined
above,
\begin{equation}
\left\langle  \chi \right\rangle_0 =-\frac{\lambda}{m}
\left\langle  \chi^2 \right\rangle_0\,.
\end{equation}
Then,
\begin{equation}
M_{\,{\rm SPM}}= \frac{1}{6} \frac{\mu_{\rm ren}^3}{\lambda^2}
- \frac{\mu_{\rm ren}}{2\pi}\,.
\label{defspm1}
\end{equation}

The definitions (\ref{defssg}) and  (\ref{defspm}) of the renormalized
mass parameter are the same as those adopted in Refs.\
\cite{NSNR,Jaffe}, and our results coincide. What then is the difference
between our calculations and those of our predecessors?

What was correctly calculated in Refs.\ \cite{NSNR,Jaffe}, as well as in
some previous works, was $\langle {\rm sol} | H |{\rm sol}\rangle $.
The problem was that the result did not match $\langle {\rm sol} | {\cal
Z}
|{\rm sol}\rangle$. What we present here is the calculation of $\langle
{\rm sol} | {\cal Z}
|{\rm sol}\rangle$. We resolve the problem by adding the anomalous term
in ${\cal Z}$.

The next section presents an effective Lagrangian approach to the
calculation of ${\cal Z}$ which reproduces the very same anomaly. 
Moreover, in SPM we suggest another definition of the renormalized
(off-shell) mass parameter, which is most natural from the point of view
of the effective Lagrangian.

\subsection{Soliton mass from effective superpotential}
\label{ssec:effsup}

A convenient notion to use is the effective superpotential
${\cal W}_{\rm eff}(\phi)$ which appears in the background field
approach to  perturbation theory.  It accounts for the quantum
corrections
in the potential energy which is considered as a function of the
background  field $\phi_{\, \rm bck} $.

Let us calculate one-loop corrections to
${\cal W}_{\rm eff}(\phi)$ choosing as the background constant
fields
$\phi$ and $F$,
\begin{equation}
\phi (x)=\phi_{\, \rm bck} +\chi (x)\,, \quad F(x) = F_{\rm bck}
+f(x)\,.
\end{equation}
 The difference with \mbox{Eq.\ (\ref{flatexp})} is that the
background
field is {\em not} a solution of any classical problem and can be
chosen
arbitrarily.
Then the part of the
Lagrangian~(\ref{minlag}) quadratic in the quantum fields
is
\begin{equation}
{\cal L}^{(2)} = \frac{1}{2}\left\{ \partial_\mu\chi\partial^\mu\chi
+f^2 +2{\cal  W}^{\prime\prime} f\chi +F {\cal
W}^{\prime\prime\prime}
\chi^2
+\bar\psi i \not\!\partial \psi
-{\cal W}^{\prime\prime}\bar\psi\psi \right\}
\, .
\end{equation}
The one-loop correction to the effective Lagrangian ${\cal L}_{\rm
eff}
(\phi_{\, \rm bck}, F_{\rm bck})$ is defined as
\begin{equation}
\exp \left\{ i \!\int {\rm d}^2 x \,\Delta{\cal L}_{\rm eff}\right\}=
\int\! {\cal D} \chi {\cal D}\psi \exp \left\{ i \!\int {\rm d}^2 x
\,{\cal
L}^{(2)}\right\}\, .
\end{equation}
As is seen from \mbox{Eq.\ (\ref{minlag})} it is sufficient to
determine  the part of $\Delta{\cal L}_{\rm eff}$ linear
in $F_{\rm bck}$ in the limit
$F_{\rm
bck} \to 0$  to get the one-loop correction to ${\cal W}^\prime_{\rm
eff}$,
\begin{equation}
\Delta{\cal W}^\prime_{\rm eff}=\frac{1}{2}  {\cal
W}^{\prime\prime\prime}
\langle \, \chi^2 \, \rangle
=\left[ \frac{{\cal W}^{\prime\prime\prime}}{8\pi}\ln
\frac{4M_r^2}{({\cal W}^{\prime\prime})^2}\right]_{\phi_{\rm
bck}}\, .
\label{derw}
\end{equation}
As mentioned above, the difference with the previous
calculation
of
$\langle\chi^2  \rangle_{0}$ is that $\phi_{\rm bck}$ is a parameter
 not equal to its classical value $\phi^+_*$.

It is easy to construct the one-loop correction to the superpotential
${\cal W}_{\rm eff}$ from its derivative~(\ref{derw}). The total
${\cal W}_{\rm eff}$  is
\begin{equation}
{\cal W}_{\rm eff}={\cal W}+\frac{{\cal
W}^{\prime\prime}}{8\pi}\left[
\ln \frac{4M_r^2}{({\cal W}^{\prime\prime})^2} +2 \right]\,.
\label{deltaw}
\end{equation}
Note the appearance of the additional 2. This is another alternative way
to reproduce the anomaly in the central charge. 
The origin of this 2 can be traced back
to the differentiation of $\ln \phi$ in the superpotential.
Thus, within the effective Lagrangian approach the anomaly emerges as
an infrared, rather than ultraviolet, effect.  This is a usual story --
both faces of the anomaly, infrared and ultraviolet, are intertwined. 

The condition ${\cal W}_{\rm eff}^\prime =0$ defines the shift of the
mean vacuum field,
\begin{equation}
\phi_{\, \rm vac}=\langle {\rm vac}|\phi | {\rm vac}\rangle =\phi^+_*
+\phi_{\,1}\,,\quad\quad
\phi_{\,1}= - \frac{1}{8\pi}\,\frac{{\cal
W}_{\,0}^{\prime\prime\prime}}{{\cal W}_{\,0}^{\prime\prime}}\,
\ln\frac{4M_r^2}{({\cal W}_{\,0}^{\prime\prime})^2}\,.
\label{meanfield}
\end{equation}
At this value of the field
\begin{equation}
{\cal W}_{\rm eff}(\phi_{\,\rm vac})={\cal
W}_{\,0} +\frac{{\cal W}_{\,0}^{\prime\prime}}{8\pi}\left[
\ln \frac{4M_r^2}{({\cal W}_{\,0}^{\prime\prime})^2} +2
\right]\,.
\label{deltawvac}
\end{equation}

The kink mass is defined by the difference between ${\cal W}_{\rm eff}$
at
$z=\pm \infty$,
\begin{equation}
M=2\,{\cal W}_{\rm eff}(\phi_{\,\rm vac})
\,,
\label{massqeff}
\end{equation}
which coincides with the result~(\ref{massqone}).
Thus, we rederive the anomaly from the effective Lagrangian.

Let us stress that the approach presented above is based heavily on
supersymmetry.  Although we did not use the superfield formalism
explicitly it was implicit  when we introduced the $F$ component of the
background field independent of $\phi$. Had we eliminated the $F$ term
from the very beginning the effective Lagrangian would differ by finite
(nonlogarithmic) terms.

The effective Lagrangian approach is also convenient for
renormalization, all ultraviolet logarithms can be hidden in
the parameters of
${\cal  W}_{\rm eff}$.
In particular, in the polynomial model
\begin{equation}
{\cal W}_{\rm eff}=\frac{m^2}{4\lambda} \phi
-\frac{\lambda}{3} \phi^3-\frac{\lambda}{4\pi} \phi
\left[\ln \frac{M_r^2}{\lambda^2 \phi^2} +2\right]\,,\quad
\phi_{\, \rm vac}=\frac{m}{2\lambda} - \frac{\lambda}{4\pi m}
\ln \frac{4M_r^2}{m^2}\;.
\end{equation}
It is natural to define the renormalized mass $m_{\rm ren}$  as
\begin{equation}
m_{\,\rm ren}=\left.- {\cal W}_{\rm eff}^{\prime\prime}
\right|_{\phi=\phi_{\rm vac}}=m - \frac{\lambda^2}{2\pi m}
\left[\ln \frac{4M_r^2}{m^2}+2\right]\;.
\label{mren}
\end{equation}
In terms of $m_{\rm ren}$ the expressions for ${\cal W}_{\rm eff}$
and $\phi_{\, \rm vac}$ contain no ultraviolet  cut off,
\begin{equation}
{\cal W}_{\rm eff}=\frac{m_{\,\rm ren}^2}{4\lambda} \phi
-\frac{\lambda}{3}\, \phi^3-\frac{\lambda}{4\pi}\, \phi
\ln \frac{m_{\,\rm ren}^2}{4\lambda^2 \phi^2} \,,
\quad \phi_{\, \rm vac}=\frac{m_{\,\rm ren}}{2\lambda}
+\frac{\lambda}{2\pi m_{\,\rm ren} }\,,
\end{equation}
and the soliton mass is
\begin{equation}
M_{\,{\rm SPM}}=\frac{1}{6} \, \frac{m_{\rm ren}^3}{\lambda^2}\;.
\label{massp4}
\end{equation}
This expression superficially looks the same as the  classical
expression. Of course, one can do things differently, expressing $M$
in terms of $\mu_{\rm ren}=2\lambda\phi_{\, \rm vac}$ introduced in
Eq.\ (\ref{defspm}).  Then we recover Eq.\ (\ref{defspm1}).
The result for the soliton mass expressed in terms of the
on-shell mass ${\overline m}$ of the elementary quanta is given in
\mbox{Eq.\ (\ref{solmonshell})}.

In the sine-Gordon model the one-loop effective potential reads as
\beq
{\cal W}_{\rm eff}= m v^2  \, \sin (\phi
/v) - {m \, \sin (\phi /v) \over 8 \, \pi} \, \left[ \ln
{4M_r^2 \over  m^2 \, \sin^2 (\phi /v)} + 2
\right
]~,
\label{weffsg}
\eeq
so that the vacuum field gets no shift \footnote{In fact, $\phi_{\rm
vac}$ stays at $\pi v /2$ to all orders.},
\beq
\phi_{\rm vac}=\phi_0= \pi \, v /2\, .
\label{vsg}
\eeq
while the renormalization of the mass parameter is given by
\beq
m_{\rm ren}=m - {m \over 8 \, \pi \, v^2} \, \ln {4M_r^2 \over
 m^2 }~.
\label{mrensg}
\eeq
This is identical to Eq.\ (\ref{defssg}).
Note that in the sine-Gordon model at one loop $m_{\rm ren}$
in \mbox{Eq.\ (\ref{mrensg})} is the same as the on-shell mass $\bar
m$ of the  elementary quanta (see Sec. \ref{sec:profile}).
The soliton mass is  the same as in \mbox{Eq.\ (\ref{ssgmass})}.

\section{Field and energy profile at one loop}
\label{sec:profile}
\setcounter{equation}{0}

\subsection{General relations for profiles}
\label{ssec:general}

The quantum corrections to the kink mass, see \mbox{Eqs.\
(\ref{massqone})}, (\ref{massqeff}), are defined by the simple dynamics
of the flat vacuum at
$z\to \pm \infty$. Here we will find the one-loop corrections to two
local
quantities: the mean field and the energy density in the kink.
We start with a general superpotential ${\cal W}(\Phi)$, and then
demonstrate the explicit solution for the polynomial and sine-Gordon
models. Similar corrections to the domain wall profile in four
dimensions were found previously in non-supersymmetric models in
\cite{mv,bs} and in supersymmetric theories in \cite{CS}.

The mean field ${\overline \phi }\,(z)$ is defined as
\begin{equation}
{\overline \phi} \,(z) =\langle {\rm sol}\, |\,  \phi (t, z) \,|\,
{\rm
sol}\rangle=
\phi_{\,0} (z) + \phi_{\,1} (z)
\;,
\end{equation}
where we denote as $\phi_{\,1} (z) $ the deviation of the mean field
from
the classical kink solution $\phi_{\,0} (z)$. To derive the equation
for the mean field let us start from the anticommutator
\begin{equation}
\{ \psi_{1}, Q_{2} \}= -\left (\partial_z  \phi + F \right )\,,
\end{equation}
where $F$ denotes $-{\cal W}^\prime (\phi)$.
As was discussed in Sec.~\ref{ssec:local} the expectation value
of the anticommutator  vanishes due to  BPS saturation,
$Q_{2}|\,{\rm sol}\rangle =0$. Therefore, we get for the mean field
\begin{equation}
\frac{{\rm d}}{{\rm d} z}\, {\overline \phi}\, (z) = -
\langle {\rm sol}\, |\,  F \,|\, {\rm sol}\rangle\, .
\label{phione}
\end{equation}
Expanding the operator field $\phi$ near the classical solution,
\begin{equation}
\phi=\phi_{\,0} (z) + \phi_{\,1} (z) +\chi (t,\,z)\;,
\end{equation}
we get from \mbox{Eq.\ (\ref{phione})}
\begin{equation}
\frac{{\rm d}}{{\rm d} z}\, \phi_{\,1} (z) = -
F^{\prime}\,\phi_{\,1} (z) -{1 \over 2} F^{''} \,\langle\,
\chi^2(z,\, t)\, \rangle_{\rm sol}\;.
\label{dfeq}
\end{equation}
Throughout this section the derivatives of the superpotential with
respect to the field $\phi$, denoted by the appropriate number of
primes, are evaluated at the classical profile, i.e. at
$\phi=\phi_0(z)$. One should also notice that the quadratic average
of
the quantum fluctuations $\langle \chi^2(t,\, z) \rangle$ is in fact
time-independent, since it is calculated in the background of a static
field configuration.

Let us first consider the behavior of the solution of
\mbox{Eq.\ (\ref{dfeq})}
at
the infinities. For the limiting behavior at $z \to \pm \infty$
we found in Sec.~\ref{sec:onemass}, see \mbox{Eq.\
(\ref{meanfield})},
\beq
\lim_{z \to \pm \infty} \phi_{\,1} (z) =  -{{\cal W}_0^{'''} \over 2 \,
{\cal W}_0^{''}}
\, \left .
\langle \chi^2(z,\, t) \rangle \right |_{z \to \pm \infty}=
 - {1 \over 8 \pi}{{\cal W}_0^{'''} \over
\,{\cal W}_0^{''}}
\,
\ln {4M_r^2 \over ({\cal W}_0^{''})^2}\bigg|_{z \to \pm \infty}~.
\label{dfinf}
\eeq
These limits are consistent with \mbox{Eq.\ (\ref{dfeq})}, its
left-hand side  is vanishing at $z \to \pm \infty$.

Using the expansion~(\ref{expansion}) of $\chi$ in eigenmodes
and
\mbox{Eq.\ (\ref{expect})} we find for the quadratic average
\begin{equation}
\langle \,\chi^2(x) \,\rangle_{\rm sol}=\sum_{n \neq 0}
\frac{\chi_n^2
(z)}{2\,\omega_n}\;.
\label{chi2}
\end{equation}
The same result   can also be presented as
\begin{equation}
\langle \,\chi^2(x) \,\rangle_{\rm sol}=G(x;x)\;,
\end{equation}
where $G(x;x)$ is the limit at coinciding space-time points of
the Green's
function $G(x; x^\prime)$,
satisfying the second order equation
\beq
\left [ -\partial_t^2 +
\left(\partial_z-F^{'}\right)\left(\partial_z+F^{'}\right)
\right ]
\, G(x; x^\prime)= i\delta(t-t^\prime)\left[ \,\delta (z-z^\prime)
- \chi_0 (z) \chi_0 (z^\prime) \right]~.
\label{gfeq}
\eeq
The subtraction of $\chi_0 (z) \chi_0 (z^\prime) $ in the right-hand
side
is needed for the exclusion of the zero mode.

Due to the time independence of the background,  the Green's function is
conveniently expanded in frequency modes:
\beq
G(t; z; t^\prime, z^\prime)= \int_{-\infty}^{+\infty} G_\omega (z;
z^\prime)
\,
e^{-i \, \omega \,
(t-t^\prime)}\, {{\rm d}\, \omega \over 2 \pi}~,
\label{gomega}
\eeq
where $G_\omega (z; z^\prime)$ satisfies the one-dimensional
equation
\beq
\left [ \omega^2 +
\left(\partial_z-F^{'}\right)\left(\partial_z+F^{'}\right)
\right ] \, G_\omega(z; z^{'})= i\, \left [ \delta(z-z^{'})- \chi_0 (z)
\chi_0 (z^\prime) \right ]~~.
\label{gomeq}
\eeq
The solution to this equation for $G_\omega (z_1;z_2)$ can be written
in
terms of the normalized eigenfunctions $\chi_n (z)$ and the
corresponding
eigenvalues $\omega_n$ defined in \mbox{Eq.\ (\ref{eigen})}
in the standard way,
\beq
G_\omega(z; z^{'})= i \sum_{n\neq 0} {\chi_n(z) \, \chi_n (z^\prime)
\over
\omega^2 -
\omega_n^2 +i\varepsilon}~~.
\label{geig12}
\eeq
At coinciding points we get
\begin{equation}
\langle \,\chi^2(x) \,\rangle_{\rm sol}=G(x;x)=
\int_{-\infty}^{+\infty}
G_\omega(z;z)\, {{\rm d}\omega \over
2 \, \pi}~,
\label{chi2prop}
\end{equation}
and the integration over $\omega$ reproduces the
representation~(\ref{chi2}).

In terms of the quantity $\langle \chi^2 (z) \rangle $ the
solution to the equation~(\ref{dfeq}) for the correction to the field
profile can be written as
\beq
\phi_{\,1} (z) = -{1 \over 2}  \chi_0 (z) \! \int_{0}^z {
F^{''}(\phi_{\,0}(u)) \,
\langle \chi^2(u) \rangle \over \chi_0(u)} \, {\rm d}u= -
{1 \over 2}  F(\phi_{\,0}(z))  \! \int_{0}^z { F^{''}(\phi_{\,0}(u)) \,
\langle \chi^2(u) \rangle \over F(\phi_{\,0}(u))} \, {\rm d}u~,
\label{dfsol}
\eeq
where  the last equality makes use of the fact that the
translational mode $\chi_0$ is proportional to $F$ due to the BPS
saturation at the classical level. The lower limit  in the integral
is the collective coordinate for the soliton center  which we have
chosen to be at $z=0$.

Once $\langle \,\chi^2(x) \,\rangle$ and $\phi_{\,1}$ are found
the quantum correction to the energy distribution in the kink
is also fully determined.  The quantity
\begin{eqnarray}
M(z)&=&2\,\left \langle\, {\cal W}\,(z)  + {{\cal W}^{''} (z) \over 4 \,
\pi}
\right \rangle_{\rm sol} =
2\,{\cal
W} +{{\cal W}^{''} \over 2 \, \pi}+ 2\,{\cal W}^\prime \,\phi_{\,1} +
{\cal W}^{\prime\prime }\,\langle \,\chi^2\, \rangle_{\rm sol}
\nonumber\\
&=& 2 \left [ {\cal W} -{F^{'} \over 4 \, \pi}-\left( F -
\frac{(F^\prime)^2}{F^{\prime\prime}}\right)\,\phi_{\,1}
+\frac{F^\prime}{F^{\prime\prime}}\frac{{\rm d} \phi_{\,1}}{{\rm d}
z}
\right ]
\label{energy}
\end{eqnarray}
represents the total energy located in the interval between $-z$ and $z$
(implying that $z>0$). In the last expression ${\cal W}$ and its
derivatives over $\phi$ are evaluated at $\phi=\phi_0 (z)$.  In the
large $z$ limit $M(z)$ goes to the soliton mass $M$. The energy density
is $(1/2) {\rm d} M (z)/{\rm d} z$.

Thus,  the one-loop corrections to the mean field and  to the energy
density are expressed via integrals over the Green's function $G_\omega
(z;z)$ defined by \mbox{Eq.\ (\ref{gomeq})}. These results were obtained
for a general superpotential function ${\cal W}$. In the special cases
of the polynomial  and sine-Gordon superpotentials (see \mbox{Eqs.\
(\ref{spot})} and (\ref{spotsg})) this Green's function is known
analytically and we will get explicit expressions for the one-loop
corrections.

\subsection{Polynomial model}
\label{ssec:spm}

The partial Green function $G_\omega(z_1;z_2)$ is readily constructed by
twice applying the relation (\ref{gfrelat}) to the free Green's function
in Eq.\ (\ref{gffree}) (it can also be read off from \mbox{Refs.\ \cite{mv,
bs}}).   At
coinciding points $G_\omega(z;z)$
is given by
\begin{eqnarray}
G_\omega(z;z)&=&-{i\over 2 \, \sqrt{ m^2-\omega^2}} \, \left [ 1 -
{1 \over \omega^2} \, {3 \, m^2 \over 4 \cosh^4 (m z/2)}
\left( 1-{
\sqrt{ m^2 -\omega^2} \over  m} \right )  \right.\nonumber\\
&& \left. + {1 \over
 3 m^2-4\omega^2} \, {3 \, m^2 \, \tanh^2 (m z/2) \over  \cosh^2 (m
z/2)}
\right ]~.
\label{gom4}
\end{eqnarray}
Thus, substituting this expression in \mbox{Eq.\ (\ref{chi2prop})} and
performing Euclidean rotation in the  integration over $\omega$ (i.e. 
the substitution $\omega \to i \,\omega$) one finds for the average over
the fluctuations
\begin{eqnarray}
\langle \chi^2(z) \rangle_{\rm sol}&&=
\int_{-\infty}^{+\infty}{d\omega \over 4 \, \pi
\,\sqrt{\omega^2 +  m^2}} \, \left [ 1 +
{1 \over \omega^2} \, {3 \, m^2 \over  4\cosh^4 (m z/2)}\, \left( 1-{
\sqrt{\omega^2+ m^2} \over m} \right )
\right.  \nonumber \\ &&
\left.+ {1 \over 4\omega^2+ 3
m^2} \, {3 \, m^2 \, \tanh^2 (m z/2)\over  \cosh^2 (m z/2)}
\right ] \label{chi2p4} \\
&&={1 \over 4 \, \pi} \, \ln {4 M_r^2 \over  m^2}
-{3 \over 8 \,
\pi} \, {1 \over \cosh^4 (m z/2)}+ {1 \over 4 \, \sqrt{3}}\,
{\tanh^2 (mz/2)
\over \cosh^2(m z/2)}~~. \nonumber
\end{eqnarray}
It is understood here that the divergent part of the integral is
regularized by the ultraviolet regularization procedure described
above.
Upon substitution in \mbox{Eq.\ (\ref{dfsol})} this expression gives
the
final result for the correction to the profile of the field across the
soliton:
\begin{eqnarray}
&&\phi_{\,1} (z)= -{\lambda^2 \over 2 \pi \, m}
\left ( \ln {4 M_r^2 \over m^2} + {2 \pi \over \sqrt{3}}
\right ) \,
\left . {\partial \phi_0 \over \partial m} \right
|_\lambda -  {\lambda^3 \over \sqrt{3} \, m^2}\left .
{\partial \phi_0 \over \partial \lambda} \right
|_m  \nonumber \\
&&+
{\lambda \over m} \, \left ( {1 \over 2 \, \sqrt{3}} + {3 \over 4 \,
\pi} \right ) {\tanh (m z/2) \over \cosh^2 (m z/2)}~.
\label{res4}
\end{eqnarray}
The first two terms in this expression clearly can be absorbed in
a redefinition of the parameters $m$ and $\lambda$ in the classical
profile of the field. Thus the full one-loop corrected field $\bar \phi$
across the soliton  can be written as
\beq
{\overline \phi} \equiv \phi_0(z)+\phi_1(z) = {{\overline m} \over
{\overline 2\lambda}} \tanh ({\overline m} z/2) +{{\overline
\lambda}
\over {\overline m}} \, \left ( {1 \over 2 \, \sqrt{3}} + {3 \over 4 \,
\pi} \right ) {\tanh ({\overline m} z /2)\over \cosh^2 ({\overline m}
z/2)}~,
\label{resfin4}
\eeq
where
\beq
{\overline m} = m - {\lambda^2 \over 2 \pi \, m}
\left ( \ln {4 M_r^2 \over  m^2} + {2 \pi \over \sqrt{3}}
\right )~~{\rm and}~~
{\overline \lambda} = \lambda - {\lambda^3 \over \sqrt{3} \, m^2}
\label{barml}
\eeq
are the appropriately renormalized parameters. It should be noted that
the mass parameter in this renormalization scheme exactly corresponds to
the on-shell mass renormalization. In other words, the physical (pole)
renormalized mass of the scalar boson (and, by supersymmetry, of the
fermion) is equal to $ \overline m$, as can be seen from the exponential
approach of the profile in \mbox{Eq.\ (\ref{resfin4})} to the vacuum
value at $z \to \infty$.

Substituting the result~(\ref{res4}) for the profile correction into
\mbox{Eq.\ (\ref{energy})} we find the energy $M(z)$ sitting
between
$-z$ and $z$,
\begin{eqnarray}
&&M(z)={{\overline m}^3 \over
6\,{\overline \lambda}^2} \tanh ({\overline m} z /2)
\left(2+\frac{1}{\cosh^2
{(\overline m} z/2)}\right)\left(1+\frac{1}{\sqrt{3}}\,\frac{
{\overline
\lambda}^2}{ {\overline m}^2}\right) \\ 
&&-{{\overline m} \over 2 \, \pi} \, \tanh {({\overline m} z/2)} +
{\overline m}\,
\frac{\tanh {(\overline m} z/2)}{\cosh^2 {(\overline m} z/2)}\left[
-\frac{1}{4\sqrt{3}} +\left(\frac{1}{2\sqrt{3}}+\frac{3}{4\pi}\right )
\frac{1}{\cosh^2  {(\overline m} z/2)} \right]
\;. \nonumber
\label{massdist}
\end{eqnarray}
The first line is similar to the classical energy distribution up to
the transition to renormalized parameters. The correction factor in this
line
is equivalent to an additional finite renormalization of $\lambda$.

The qualitative feature which quantum corrections bring in is that
$M(z)$ approaches its asymptotic value at large $z$ more slowly.
Classically (the first line) it approaches as $\exp (-2 \,  m z)$. The
one-loop correction contains  $\exp (-  m z)$. This can be simply
understood in perturbation theory: the energy density operator can only
produce one-particle states at the quantum level.

We also find it quite instructive to consider a
``kink effective superpotential" ${\overline {\cal W}}_{\rm
eff}({\overline
\phi})$, which generates the full one-loop profile of the field, as the
solution to an  ``effective" equation
\beq
{{\rm d} \over {\rm d} z} \,{\overline\phi}=
{\overline {\cal W}}_{\rm eff}^{\;'}({\overline \phi})~~.
\label{effeq}
\eeq
This effective function can readily be found by noticing that the
expression for the average $\langle \chi^2(z) \rangle_{\rm sol}$
given
by \mbox{Eq.\ (\ref{chi2p4})}, which enters the quantum BPS
equation
(\ref{dfeq}), is in fact a fourth power polynomial in the classical
solution $\phi_{\,0}(z)$. Thus the inhomogeneous term in
\mbox{Eq.\ (\ref{dfeq})}
can be written as the derivative of a fifth power polynomial in
$\phi$.
Hence
one arrives at the following expression for ${\overline {\cal
W}}_{\rm eff}$
\beq
{\overline {\cal W}}_{\rm eff}({\overline \phi})= {{\overline m}^2
\over
{4\overline \lambda}} \,
{\overline \phi} - {{\overline \lambda} \over 3} \, {\overline \phi}^3
+
{\overline
\lambda } \,
\left ( {1 \over 4 \, \sqrt{3}} + {3 \over 8 \,
\pi} \right ) \, \left [ {\overline \phi} - { 8 {\overline \lambda}^2
\over 3
\,  {\overline m}^2} \,
{\overline \phi}^3 + { 16 {\overline \lambda}^4 \over 5 \, {\overline
m}^4} \,
{\overline \phi}^5 \right ]~~.
\label{kweff}
\eeq

It is important to emphasize that the so defined soliton effective
(super)potential ${\overline {\cal W}}_{\rm eff}$ is {\it different}
from the Coleman-Weinberg type effective (super)potential
${\cal W}_{\rm eff}$ in \mbox{Eq.\ (\ref{deltaw})}. The latter
describes
the response of the
action to a space-time constant field, and is appropriate for finding
the quantum corrections to the vacuum field, while the purpose of
the
former is to describe the quantum corrections to the soliton profile
through the equation (\ref{effeq}). In fact, the ${\overline {\cal
W}}_{\rm eff}$ also describes the correction to the vacuum field as
the
asymptotic value of the solution at $z \to \infty$, which perfectly
agrees with $\phi_{\rm vac}$ found from ${\cal W}_{\rm eff}$:
${\overline {\cal W}}_{\rm eff}^{'}(\phi_{\rm vac})=0$. In addition
the
soliton effective superpotential encodes all information about the
detailed behavior of the field profile at finite $z$. In particular, its
second derivative gives the on-shell mass of the
elementary particles,
\beq
\left . {\overline {\cal W}}_{\rm eff}^{''}\right |_{\phi=\phi_{\rm
vac}}
= - {\overline m}\, ,
\label{polem}
\eeq
unlike the mass parameter $m_{\rm ren}$, calculated in
\mbox{Eq.\ (\ref{mren})}
from the curvature of ${\cal W}_{\rm eff}$,
\beq
{\overline m}= m_{\rm ren}+{\lambda^2 \over \pi \, m} \,
\left ( 1- {\pi \over \sqrt{3}} \right )\, .
\label{diffm}
\eeq
Thus,  the mass of the kink from \mbox{Eq.\ (\ref{massp4})}, being
expressed
through the on-shell mass parameter ${\overline m}$ reads as
\beq
M={1 \over 6}\, {{\overline m}^3 \over \lambda^2}+
{{\overline m} \over 2\sqrt{3}}- {{\overline m} \over 2 \, \pi}=
{1\over
6}\, {{\overline m}^3 \over
{\overline \lambda}^2}+{{\overline m} \over 6 \sqrt{3}} - {{\overline
m}
\over 2 \, \pi}
\, .
\label{solmonshell}
\eeq
The same expression for $M$ appears in \mbox{Eq.\
(\ref{massdist})}
as the asymptotic value of $M(z)$ at large $z$. The first two terms in
(\ref{solmonshell}) are equivalent to the result of \mbox{Ref.\
\cite{HY}}
where it was obtained within a similar approach. However our result
differs by the third term, which arises from the anomalous part of
the
central charge.

A remarkable property of the soliton effective (super)potential is that
it is a finite polynomial of the field, unlike that of the
Coleman-Weinberg type. This is due to the phenomenon of ``nullification''
~\cite{mv,bs,mv1}: the vanishing of all on-shell amplitudes for
multiboson production at threshold, starting from a finite number of the
produced bosons. The on-shell amplitudes are related to the profile of
the soliton field~\cite{brown,mv}.

\subsection{Sine-Gordon model}
\label{ssec:ssg}

Following the same routine as for the polynomial model, we write the
expression for the partial Green's function $G_\omega(z; \, z)$ in the
background of the classical soliton profile given by \mbox{Eq.\
(\ref{thsg})},
\beq
G_\omega(z; \, z)= - {i \over 2 \, \sqrt{ m^2 - \omega^2}} \left [
1- {1 \over \omega^2} \, { m^2  \over \cosh^2  m  z} \, \left ( 1-
{\sqrt{ m^2 - \omega^2} \over  m} \right ) \right ]~.
\label{gozzsg}
\eeq
The average over the quantum fluctuations then takes the form
\begin{eqnarray}
&&\langle \chi^2(z) \rangle_{\rm sol}=
\int_{-\infty}^{+\infty}{d\omega \over 4 \, \pi
\,\sqrt{\omega^2 +
 m^2}} \, \left [ 1 +
{1 \over \omega^2} \, { m^2 \over  \cosh^2  m z}\, \left( 1-{
\sqrt{\omega^2+ m^2} \over  m} \right )
\right ] \nonumber \\
&&={1 \over 4 \, \pi} \, \ln {4 M_r^2 \over  m^2} - {1 \over
2\, \pi} \, {1 \over \cosh^2  m z}~.
\label{chi2sg}
\end{eqnarray}
Thus, according to \mbox{Eq.\ (\ref{dfsol})}, the one-loop correction
to
the soliton profile is given by
\beq
\phi_1(z)=-{m \over 8 \, \pi \, v^2} \, \ln {4 M_r^2 \over
m^2} \, \left . {\partial \phi_0 \over \partial m} \right |_v + {1 \over
4 \, \pi \, v} {\tanh m z \over \cosh m z}~.
\label{phi1sg}
\eeq
Absorbing the first term in this expression into a renormalization of
the
mass in the classical field profile, one can write the final expression
for the full one-loop profile of the soliton field ${\overline \phi}$ as
\beq
{\overline \phi}(z)=2 \, v \, \arctan [ \tanh ({\overline m} z/2)] +
{1 \over 4 \, \pi \, v} {\tanh {\overline m} z \over \cosh
{\overline m} z}~,
\label{phibarsg}
\eeq
where
\beq
{\overline m}=m - {m \over 8 \,  \pi \, v^2} \, \ln {4 M_r^2
\over
 m^2}
\label{mbarsg}
\eeq
is the renormalized on-shell mass of the elementary particles in the
sine-Gordon model.
One can readily notice that in this model at one loop the
renormalized
pole mass coincides with the one defined from the one-loop effective
potential: $m_{\rm ren}$ in \mbox{Eq.\ (\ref{mrensg})}. This is  no
surprise, since in this model at one loop there is no dispersion of the
particles' self-energy. (This property, and the relation ${\overline
m}=m_{\rm ren}$ is  lost, however, at the two-loop level, see 
Sec.\ref{2loops}.)

The amount of energy located in the interval between $-z$ and $z$
is
readily calculated from \mbox{Eq.\ (\ref{energy})} and is given by
\beq
M(z)= 2 \, {\overline m} \, \left( v^2 - {1 \over 4 \, \pi} \right ) \,
\tanh
{\overline m} z + {
{\overline m} \over \pi} \, {\tanh  {\overline m} z \over \cosh^2
{\overline m} z}~.
\label{massdistsg}
\eeq
One can notice that in this case the approach of $M(z)$ to its
asymptotic value $M$ is described by terms proportional to $\exp(-2
mz)$, i.e. corresponding to propagation of a two-particle state. This is
due to the fact that in the sine-Gordon model the  number
of particles is conserved modulo 2, so that the energy operator
does not couple
through quantum loops to a one-particle state.

The soliton effective superpotential is also readily recovered in this
model and reads as follows,
\beq
{\overline {\cal W}}_{\rm eff}= {\overline m} \, v^2 \, \sin
{{\overline \phi} \over v} + {{\overline m} \over 4 \, \pi} \left ( \sin
{{\overline \phi} \over v} - {1 \over 3} \, \sin^3 {{\overline \phi}
\over v} \right )~,
\label{wbarsg}
\eeq
and is polynomial in $\sin\, ({\overline \phi} / v)$.

\section{ ``Supersymmetry'' of the non-SUSY
sine-Gordon model. }
\label{sec:recurrency}
\setcounter{equation}{0}

Here we will discuss a ``supersymmetry" in the problem of
calculating the soliton mass (at one loop) in the
 non-supersymmetric sine-Gordon
model, whose Hamiltonian is simply the bosonic part of that of the
SSG
model. The sine-Gordon Hamiltonian density can be
written as
\beq
{\cal H}= \partial_z {\cal W} + {1 \over 2} \left \{ \dot \phi^2 + \left
( \partial_z \phi +F \right )^2 \right \}
\label{sgham}
\eeq
with ${\cal W(\phi)}$ given by Eq. (\ref{spotsg}) and $F(\phi)=-{\cal
W(\phi)}^{'} = -m \, v \, \cos(\phi/v)$.
Since the purpose of this section is mainly illustrative, we will ignore
the necessity for ultraviolet regularization. This can be done along
the lines of Sec. \ref{ssec:cahdr}. We will not introduce the explicit
regularization  to avoid cumbersome formulae which could
overshadow our key points. One should understand that in doing so
we are bound to miss some finite (non-logarithmic) piece in the soliton
mass. This flaw could easily be eliminated, if desired.

In order to write the one-loop (quadratic) expression for the energy in
terms of the quantum fluctuations $\chi$ over the kink background
$\phi_0(z)$ (see  Eq. (\ref{thsg})) one should  take into account that,
unlike  the supersymmetric case, the flat vacuum energy density
receives a one-loop renormalization and is to be subtracted. The
proper expression is as follows:
\beq
\langle {\cal H}\rangle_{\rm kink} - \langle {\cal H} \rangle_{\rm
vac}-
\partial_z {\cal W}  =
{1 \over 2} \left \{ \dot \chi^2 + \left [\left ( \partial_z  +F^{'}
\right ) \, \chi\right ]^2 - \dot {\tilde \chi}^2
- \left ( \partial_z {\tilde \chi} \right )^2 - m^2 {\tilde \chi}^2
\right \}~.
\label{sgham2}
\eeq
Here  $F^{'}= m \, \tanh mz$ is the background value at
$\phi_0(z)$, and ${\cal W}$ is the profile of $m \, v^2 \,
\sin(\phi/v)$ across the soliton  (including the one-loop correction).
Finally,  ${\tilde \chi}$ denotes in this section the field
fluctuations over the flat vacuum at $\phi_0=\pi v/2$
(see Sec. \ref{ssec:irreg}).

We consider the contribution of the quantum fluctuations $\chi$
and $\tilde \chi$, explicitly written
in Eq. (\ref{sgham2}), by expanding these fields in the appropriate
eigenmodes. Namely, the field $\chi$ is expanded
in the eigenmodes $\chi_n(z)$ of the operator
\beq
L_2=-\left (\partial_z-F^{'} \right )\, \left (\partial_z+F^{'} \right
)=
-\partial_z^2+ m^2 - {2 \, m^2 \over \cosh^2 m z}~,
\label{lsg}
\eeq
while the corresponding operator for the modes ${\tilde \chi}_n$  of the
field ${\tilde \chi}$ in the flat vacuum is obviously
\beq
{\tilde L}_2=-\partial_z^2+ m^2~.
\label{lflat}
\eeq

Since the operators $L_2$ and ${\tilde L}_2$ are related by a
quantum-mechanical  supersymmetry (see Sec.~\ref{ssec:irreg}), we use
here the fact that the spectra of both operators coincide (modulo the
zero mode of $L_2$). We also use the ensuing relations for the
eigenmodes. It is seen that the soliton sector can be  dubbed ``bosonic"
while the flat vacuum ``fermionic". These two sectors  are related by a
quantum-mechanical  ``supercharge" $P=(\partial_z + F^{'})$.

With our boundary conditions, the terms in Eq. (\ref{sgham2}) can be
formally~\footnote{As was mentioned, for illustrative purposes we
ignore
here the problem of the specific ultraviolet regularization of the
one-loop terms in the sine-Gordon model. Thus, an analog of 
the anomalous term
${\cal W}''/4\pi$ in the central charge will be lost. We will focus only
on the
logarithmically divergent part of the soliton mass renormalization.}
written as follows:
\begin{eqnarray}
&&\left \langle \dot \chi^2 \right \rangle = \sum_{\omega_n \neq0}
{\omega_n \over 2} \chi_n^2(z)~, \nonumber \\
&&\left \langle \left [\left ( \partial_z  +F^{'}
\right ) \, \chi\right ]^2 \right \rangle = \sum_{\omega_n \neq0}
{\omega_n \over 2} {\tilde \chi}_n^2(z)~, \nonumber \\
&&\left \langle \dot {\tilde \chi}^2 \right \rangle = \sum_{\omega_n
\neq0} {\omega_n \over 2} {\tilde \chi}_n^2(z)~, \nonumber \\
&&\left \langle \left ( \partial_z {\tilde \chi} \right )^2 + m^2
{\tilde \chi}^2 \right \rangle={1 \over 2}\partial_z^2 \,
\sum_{\omega_n
\neq0}{{\tilde \chi}_n^2(z) \over 2 \omega_n}+ \sum_{\omega_n
\neq0}{\omega_n \over 2} {\tilde \chi}_n^2(z)~.
\label{sgaver}
\end{eqnarray}
Here the relation
$$P \, \chi_n(z)= \omega_n
\, {\tilde \chi}_n(z)$$
 is used  in the second line. Note also that the total derivatives are
explicitly kept. Using these relations, the expression (\ref{sgham2})
for
the kink energy density can be formally rearranged as
\beq
\langle {\cal H}\rangle_{\rm kink} - \langle {\cal H} \rangle_{\rm
vac}-
\partial_z {\cal W} =-{1 \over 4}\partial_z^2 \, \sum_{\omega_n
\neq0}{{\tilde \chi}_n^2(z) \over 2 \omega_n}+{1 \over 2}
\sum_{\omega_n
\neq 0} {\omega_n \over 2} \left [ \chi_n^2(z)-{\tilde \chi}_n^2(z)
\right ] ~.
\label{sgsum}
\eeq

It can be noticed, that the integral of each term in the last sum in the
right-hand side of the latter equation over the whole box  vanishes
identically, since the modes in both sectors are normalized to one.
However the relevant integral over the fiducial interval, discussed in
the Sec.\ \ref{ssec:separation} is not zero, due to different
behavior of the modes at the edges of the bounding box\footnote{A
similar behavior can be seen from Fig.\ \ref{fig:chi2}: the integrals of
the two curves over the whole box are equal, while the integrals over
the fiducial interval are clearly different.}. In order to calculate the
integral over the latter interval we again use
the quantum-mechanical supersymmetry to express the modes $\chi_n$ in
terms of ${\tilde \chi}_n$,
\begin{eqnarray}
&&\chi_n^2={1 \over \omega_n^2}\left( P^\dagger {\tilde \chi}_n
\right)^2 ={1 \over \omega_n^2} \partial_z \left [  {\tilde \chi}_n
\left(
\partial_z - F^{'} \right) {\tilde \chi}_n \right ] + {{\tilde \chi}_n
P P^\dagger {\tilde \chi}_n \over \omega_n^2}= \nonumber \\
&&{1 \over \omega_n^2} \partial_z \left[  {\tilde \chi}_n \left(
\partial_z - F^{'} \right) {\tilde \chi}_n \right]+ {\tilde
\chi}_n^2~.
\label{chitchi}
\end{eqnarray}
Then   Eq. (\ref{sgsum}) can be rewritten as
\beq
\langle {\cal H}\rangle_{\rm kink} - \langle {\cal H} \rangle_{\rm
vac} -
\partial_z \left[ {\cal W} -
{1 \over 2} \,
F^{'} \, \sum_{\omega_n \neq0} { {\tilde \chi}_n^2 (z)
\over 2 \omega_n} \right]
= {1 \over 4}\partial_z^2 \, \sum_{\omega_n
\neq0}{{\tilde \chi}_n^2(z) \over 2 \omega_n}\, .
\label{sgsum2}
\eeq

The soliton mass is given by  the integral over the fiducial interval,
i.e. over $z$ from $-L/2+\Delta$
to $+L/2-\Delta$ where $\Delta$ is chosen
in such a way that $L\gg\Delta \gg m^{-1}$ (see the discussion in
Sec.\ \ref{ssec:separation}).
Then it is not difficult to see that the right-hand side in Eq.(\ref{sgsum2})
vanishes identically within the fiducial interval, and
the soliton mass is given by the following formula
\beq
M =
\left[ {\cal W} -
{1 \over 2} \,
F^{'} \, \sum_{\omega_n \neq0} { {\tilde \chi}_n^2 (z)
\over 2 \omega_n} \right]_{L/2 - \Delta}-
\left[ {\cal W} -
{1 \over 2} \,
F^{'} \, \sum_{\omega_n \neq0} { {\tilde \chi}_n^2 (z)
\over 2 \omega_n} \right]_{-L/2 + \Delta}\, .
\eeq
This result is perfectly analogous to what we had in the SSG model,
except that the expression for the ``central charge" changes,
the one-loop correction to the soliton mass is given by
\beq
\Delta M = \Delta\langle  {\cal W}'' \chi^2 \rangle\,  ,
\eeq
with the coefficient of $\langle \chi^2 \rangle $ twice larger than it
would be in a calculation of the correction to
$\langle {\cal W} \rangle$ alone. (Note though that the classical part
of $M$ is determined by
$\Delta{\cal W}$.) As a result,
\beq
M=2 \left[ mv^2-\frac{m}{4\pi}\, \ln\frac{4M_r^2}{m^2}\right] \,  .
\label{nsg}
\eeq

This properly  reproduces the part of the one-loop correction to the
kink mass logarithmic in $M_r$ in the non-supersymmetric sine-Gordon
model. The finite part is lost in the above reasoning because of our
formal manipulations with divergent sums. This can be readily fixed by
considering explicitly an ultraviolet regularization. The finite
contribution to the kink mass in fact is well known (see e.g. in
\cite{rajaraman}) and is equal to $-m/(2 \pi)$. We do not present
details here, since the goal of this section is to illustrate the
implementation of the supersymmetric (in the sense of Witten's quantum
mechanics) boundary conditions and the separation  of the kink energy
from the energy of the boundaries.

\section{Anomaly and soliton mass in the SSG model at two loops}
\label{2loops}
\setcounter{equation}{0}

We have shown in Sec.\ \ref{ssec:highloops} that the anomaly relation
(\ref{anocc}) is not corrected beyond one loop. In the framework of the 
SSG model we now present two arguments confirming this statement.

The first is based on the relation for the superpotential in the SSG
model
\beq
\left \langle {\cal W}^{\prime\prime}(\phi) \right\rangle =
-\frac{1}{v^2}\left \langle {\cal W} (\phi)\right\rangle \,,
\label{w2pw}
\eeq
which holds at all orders of perturbation theory. Using this relation
in the expression for the anomaly, one finds, assuming that the
one-loop
expression for the anomaly is exact, that at
$$
v^2={1 \over 4 \, \pi}
$$
the central charge and also $\vartheta_\mu^\mu$ vanish and, thus, the
theory
becomes conformally invariant, which is the known behavior of the SSG
model at this value of the coupling (see e.g. \cite{Ahn}).

The second confirmation comes from an explicit two-loop
calculation of the kink mass, presented below,  and its comparison with
the known result \cite{Ahn} found under the assumption of
factorizability of the $S$ matrix in this theory.  The calculation of
the kink mass in the SSG model at the multi-loop level was most recently
considered in \mbox{Ref.\ \cite{NSNR}}, where an actual two-loop
calculation was also done in detail. We believe that our approach, based
on the anomalous relation (\ref{anocc}), provides us with a much simpler
way of calculating the soliton mass, at least at the two-loop level. The
calculation is presented in this section.

\begin{figure}[h]
  \begin{center}
    \leavevmode
    \epsfbox{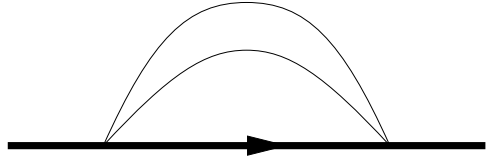}
    \caption{Non-tadpole two loop graph in the fermion self-energy.
         The heavy line denotes the fermion, and the thin lines denote 
         bosons.}
    \label{fig:2loop}
  \end{center}
\end{figure}

According to  Eq.\ (\ref{anocc}) the mass of the soliton in the
SSG model is given by
\beq
M=2 \left \langle{\cal W}(\phi) + {{\cal W}^{\prime\prime}(\phi) \over 4
\pi}
\right
\rangle_{\pi v/2} = 2 \,  \left ( 1- {1 \over 4 \pi \, v^2} \right ) \,
\left \langle {\cal W} (\phi) \right \rangle_{\pi v/2} ~~,
\label{m2l}
\eeq
where the subscript $ \pi v/2$ denotes that the quantum average of the
superpotential is taken over the ``flat'' vacuum at $\langle \phi
\rangle = \pi v /2$. The relation (\ref{w2pw}) is also taken into
account in \mbox{Eq.\ (\ref{m2l})}. Although the calculation of the
quantum average of the superpotential function in two (or higher) loops
is quite straightforward in any given regularization scheme in terms of
the regulator parameter $M_r$ and the bare mass $m$, we prefer not to
present it explicitly, since this part is completely absorbed in the
renormalization of the particles' mass. Technically the reasoning is most
straightforward for the on-shell fermion mass, which at the two loop
level can be
written as
\beq
m_F=-\left\langle {\cal W}^{\prime\prime}(\phi) \right\rangle_{\pi v/2}
+
\delta_2 m={M
\over 2
\,
\left ( v^2- {1 \over 4 \pi } \right )} + \delta_2 m~~,
\label{mf2}
\eeq
where the equations (\ref{m2l}) and (\ref{w2pw}) are used in the
last equality, and $\delta_2 m$ is the non-tadpole contribution to the
fermion self-energy given by a single graph of
\mbox{Fig.\ \ref{fig:2loop}}.
The tadpole part
of the mass, given by $\langle {\cal W}^{''}(\phi) \rangle$ has no
dispersion,
while the finite graph of Fig.\ \ref{fig:2loop} introduces dispersion
of the self-energy
in fermion momentum $p$. For the on-shell mass of the fermion, we
calculate this graph at $\not \! p = m$ and find:
\beq
\delta_2m=-{m^2 \over 2 \, v^4} \left. \int \!{{\rm d}^2 k \over
(2\pi)^2}
\, {{\rm d}^2 q \over (2\pi)^2} \, {\not\! p + \not \! q +m \over
(p+q)^2-m^2} \, {1 \over (k^2-m^2)} \, {1 \over (q+k)^2-m^2 }
\right|_{\not \, p=m}=-{m
\over 96 v^4}~~.
\label{d2m}
\eeq

Thus, up to the two-loop terms, inclusively, (i.e. of the relative
order $v^{-4}$) the mass of the kink is related to the renormalized
on-shell mass $\overline m$ of the particles in the SSG model as
\beq
M_{\rm two-loop}= 2 \, {\overline m} \, \left( v^2 - {1 \over 4 \,
    \pi} \right)+ {{\overline m} \over 48 \, v^2}~~.
\label{m2loop}
\eeq
This relation is  in perfect agreement with the expansion, up to the
appropriate order ($\gamma^3$), of the known exact relation
\beq
{\overline m} = 2 \, M \, \sin {\gamma \over 16}
\label{yb}
\eeq
with
$${4 \over \gamma} = {v^2 } - {1 \over 4 \, \pi}~,$$
found \cite{Ahn} within the assumption of a factorized $S$ matrix in
the
SSG model.

\section{Extra superfields and extended supersymmetry}
\label{sec:extended}
\setcounter{equation}{0}

In this section we will consider
the ${\cal N}=2$ model in two dimensions, obtained by
dimensional reduction of the four-dimensional Wess-Zumino model
\cite{WZ}, and its deformations which break ${\cal
N}=2$ down  to ${\cal N}=1$. A general classification of ${\cal N}=2$ theories in two
dimensions is given in~\cite{Vafa}.

In four dimensions the Wess-Zumino model realizes minimal, ${\cal N}=1$,
supersymmetry with four fermionic generators.  In two dimensions it
means that an extra pair of fermionic generators is added to
superalgebra~(\ref{ce}). Let us briefly recall this construction putting
emphasis on those elements that are specific for ${\cal N}=2$.

The Wess-Zumino model in four dimensions
describes (self-)interactions of one {\em complex} scalar field
$\varphi
(x)$
 and one {\em complex} Weyl spinor
$\psi^\alpha
(x)\, , \,\,\, \alpha = 1,2$ .
Both the boson and fermion degrees of freedom are united in one
{\em chiral} superfield \footnote{For our conventions and notation
see Appendix A in \mbox{Ref.\ \cite{CS}}}.
\beq
{\Phi ({x}_L,\theta )} = \varphi ({x}_L) + \sqrt{2}\theta^\alpha
\psi_\alpha ({
x}_L) +
\theta^2 f({x}_L)\, ,
\label{chsup}
\eeq
where $\theta$ is a two-component complex Grassmann spinor. The
chiral coordinate $x_L$ is defined as
\beq
({x_{L}})_{\alpha\dot{\alpha}} = {x}_{\alpha\dot{\alpha}} -
2i\, \theta_{\alpha}\bar{\theta}_{\dot{\alpha}}\, ,
\label{chcoor}
\eeq
so that the supertransformation takes the form
\beq
\theta_\alpha \rightarrow\theta_\alpha +\varepsilon_\alpha\, ,
\,\,\,  ({ x}_L)_{\alpha\dot\beta} \rightarrow
({x}_L)_{\alpha\dot\beta} -
4i\, \theta_{\alpha}\bar\varepsilon_{\dot\beta}\, ,
\eeq
where $\alpha\, , \,\, \dot\alpha = 1,2 $ are the  Lorentz indices.
The Lagrangian of the Wess-Zumino model in terms of superfields is
\beq
{\cal L}_{{\rm WZ}} = \frac{1}{4}\int d^4\theta \, \Phi\bar{\Phi} +
\left\{ \frac{1}{2}\int d^2 \theta\, {\cal W}_{{\rm WZ}}(\Phi) +
H.c.\right\}\, ,
\label{lagrwz}
\eeq
where $ {\cal W}_{{\rm WZ}}$ is a superpotential. We choose it to be
of the form
\beq
{\cal W}_{{\rm WZ}}(\Phi) =
\frac{i}{\sqrt{2}}\left[\frac{m^2}{4\lambda} \,\Phi -
\frac{2\lambda}{3} \Phi^3 \right]
\label{spot4}
\,,
\eeq
where $m$ and $\lambda$ are real parameters.

In components the Lagrangian takes the form
\beq
{\cal L}_{{\rm WZ}} = \partial^{\mu} \varphi^+\partial_{\mu} \varphi
+
\psi^{\alpha} i \partial_{\alpha\dot\alpha}\bar{\psi}^{\dot\alpha} +
f^+f
 + \left\{ f\,{\cal W}_{{\rm WZ}}^\prime (\varphi) -  \frac{1}{2} {\cal
W}_{{\rm
WZ}}^{\prime\prime}(\varphi )\psi^\alpha\psi_\alpha
  +{\rm H.c.}\right\} \, ,
\label{wzcomp}
\eeq
where the auxiliary field $f$ can be eliminated by
virtue of the classical equation of motion, $ f^+ = - {\cal W}_{{\rm
WZ}}^\prime$.

Now, to perform the reduction to two dimensions it is sufficient to
assume that all fields are independent of $x$ and $y$, they depend on
$t$ and $z$ only. In two dimensions the Lagrangian can be presented in
the following form:
\begin{equation}
{\cal L}=\frac{1}{2}\left\{ \partial _\mu\varphi_i
\,\partial^\mu\varphi_i
+i \bar \psi_i \gamma^\mu \partial_\mu \psi_i + f_i f_i +2 f_i
\,\frac{\partial
{\cal W}}{\partial \varphi_i} - \frac{\partial^2 {\cal W}}{\partial
\varphi_i \partial
\varphi_j}\, \bar
\psi_i
\psi_j \right\}
\;.
\label{n2lag}
\end{equation}
We introduced real fields $\varphi_i$, $\psi_i$ and $f_i$, where
$i=1,2$.  Up to
normalization they are just real and imaginary parts  of the original
fields, e.g,
$\varphi=(\varphi_1+i \varphi_2)/\sqrt{2}$. Summation over
$i$ is implied.  The superpotential ${\cal W}=2\, {\rm Im} {\cal
W}_{{\rm WZ}} $ is
a function of two variables
$\varphi_1$ and $\varphi_2$,
\begin{equation}
{\cal W}(\varphi_1, \varphi_2) = \frac{m^2}{4\lambda} \,\varphi_1 -
\frac{\lambda}{3} \,\varphi_1^3 +\lambda\,\varphi_1\varphi_2^2~.
\end{equation}

The Lagrangian can also be presented in terms of ${\cal N}=1$
superfields $\Phi_i$ (see \mbox{Eq.\ (\ref{n1super})}  for the
definition), namely
\begin{equation}
{\cal L} =i\!\int\! {\rm d}^2\theta\,  \left\{ \frac{1}{4}\bar D_\alpha
\Phi_i
D_\alpha\Phi_i +{\cal W}(\Phi_1, \Phi_2 )\right\}\;.
\label{minac2}
\end{equation}
The presence  of extended supersymmetry is reflected in the
harmonicity
of this  superpotential,
\begin{equation}
\frac{\partial^2 {\cal W}}{\partial \varphi_i \partial
\varphi_i}=0\qquad \mbox{for ${\cal N}=2$}
\,.
\end{equation}

We will consider  a more general case of nonharmonic ${\cal
W}\,(\varphi_1,
\varphi_2) $,
\begin{equation}
{\cal W}\,(\varphi_1, \varphi_2) = \frac{m^2}{4\lambda} \,\varphi_1
-
\frac{\lambda}{3} \,\varphi_1^3 +q\,\lambda\,\varphi_1\phi_2^2
+\frac{p\,m}{2}\,\varphi_2^2\,,
\end{equation}
where
$p$ and $q$ are dimensionless parameters.
For $q\neq 1,\; p\neq 0$ the extended ${\cal N}=2$ supersymmetry
is explicitly broken down
to ${\cal N}=1$.

The classical solution for the kink is the same as in the
SPM
model with one superfield considered earlier; the second field
$\varphi_2$  vanishes,
\begin{equation}
\varphi_1= \phi_0 (z)=\frac{m}{2\lambda}\tanh \frac{mz}{2}\,,
\quad
\varphi_2=0~.
\label{N2sol}
\end{equation}

The topological current $\zeta^\mu$ now takes the form,
\begin{equation}
\zeta^\mu=\epsilon^{\mu\nu}\partial_\nu \left[ {\cal W} +
\frac{1}{4\pi}\frac{\partial^2 {\cal W}}{\partial \varphi_i \partial
\varphi_i}\right]\,.
\end{equation}
For the harmonic superpotentials, i.e. ${\cal N}=2$, the anomaly
vanishes.

The quadratic expansion of the energy density ${\cal H}$ around  the
solution (\ref{N2sol})
looks as follows (compare with \mbox{Eq.\ (\ref{H2})}):
\begin{eqnarray}
\left[{\cal H}-\zeta^{\,0}\right]_{\rm quad}
=\mbox{\hspace{8.6cm}}&&
\nonumber\\
\frac{1}{2}\left\{ \dot \chi_1^2
+[(\partial_{\,z} +2\lambda\phi_{\,0})\chi_1]^2
+i\,\eta_1(\partial_{\,z} -
2\lambda\phi_{\,0})\xi_1 +i\, \xi_1(\partial_{\,z}
+2\lambda\phi_{\,0})\eta_1 \right. &&\nonumber\\
\left. +\dot \chi_2^2
+[(\partial_{\,z} -2q\lambda\phi_{\,0}-p\,m)\chi_2]^2
+i\,\eta_2(\partial_{\,z} +
2q\lambda\phi_{\,0}+p\,m)\xi_2 \right.
\mbox{\hspace{0.3cm}} &&\nonumber\\
\left.
+i\, \xi_2(\partial_{\,z}
-2q\lambda\phi_{\,0}-p\,m)\eta_2
\right\}\,,\mbox{\hspace{5.8cm}}&&
\label{H22}
\end{eqnarray}
where the following notation is used:
\begin{equation}
\chi_1=\varphi_1-\phi_{\,0}\,,\quad \chi_2=\varphi_2\,, \quad
\left(
\begin{array}{l}
\xi_i
\\
\eta_i
\end{array}\right)
=\psi_i
\;.
\end{equation}
The part of this expression containing the fields marked by the
subscript  1 is exactly the same as in
\mbox{Eq.\ (\ref{H2})}; the  fields with the subscript 2  are new.

What changes do these new fields introduce? First of all,  there is a
new fermionic zero mode in the field $\xi_2$, \begin{equation}
\xi_2^{\,{\rm zero}}= \frac{\exp\,(-p\, mz)}{\cosh^{2\, q}\,(mz/2)}\;.
\end{equation} At $p=0,~q=1$ it has the same functional form as the old
fermionic mode in $\eta_{\,1}$. This is not surprising  because  ${\cal
N}=2$ supersymmetry places the kink in the supermultiplet which contains
two bosonic and two fermionic solitons (the second bosonic kink
corresponds to both fermionic modes filled in). What is more intriguing
is that the same degeneracy remains after ${\cal N}=2$ breaking. If
$p\neq 0$ and/or $q\neq 1$ the ${\cal N}=2$ SUSY is replaced by ${\cal
N}=1$. Still, two fermion zero modes are present. The Jackiw-Rebbi index
\cite{Jackiw} provides a general argument proving  the existence of the
zero mode in  $\xi_2$. Thus, we have two degenerate fermionic solitons,
and two bosonic. The degeneracy is true only for the ground solitonic
states. The ${\cal N}=2$ degeneracy of the nonzero modes is lifted at 
$p\neq 0$ and/or $q\neq 1$.

For the  nonzero modes of $\chi_2$, $\xi_2$ and $\eta_2$ two new
differential
operators come into play,
\begin{eqnarray}
-(\partial_{\,z} -2q\lambda\phi_{\,0}-p\,m)(\partial_{\,z}
+2q\lambda\phi_{\,0}+p\,m) v_n(z) = {\tilde \omega}_n^2
v_n(z)\,,&&
\nonumber\\
-(\partial_{\,z} +2q\lambda\phi_{\,0}+p\,m)(\partial_{\,z}
-2q\lambda\phi_{\,0}-p\,m){\tilde v}_n(z) = {\tilde
\omega}_n^2{\tilde
v}_n(z)\,,&&
\label{neweigen}
\end{eqnarray}
Again, at $p=0,~q=1$ they coincide with the operators of
\mbox{Eqs.\ (\ref{eigen})} and
(\ref{eigen2}),  due to ${\cal N}=2$. This degeneracy is lifted in the
general case,
but  the degeneracy of the eigenvalues of the
operators~(\ref{neweigen}) stays
intact.
Their eigenmodes are related,
\begin{equation}
{\tilde v}_n(z) =\frac{1}{{\tilde
\omega}_n} (\partial_{\,z}+2q\lambda\phi_{\,0}+p\,m)v_n(z)
\end{equation}
Expanding $\xi_2$ in $v_n(z)$, $\chi_2$, $\eta_2$ in ${\tilde
v}_n(z) $
and quantizing the coefficients in the same way as in Sec.
\ref{ssec:local}  we observe the explicit fermion-boson
cancellation
in the expectation value over the soliton state,
\begin{equation}
\left \langle \,{\rm sol}|\,{\cal H}(x)-\zeta^{\,0}(x)|\,{\rm
sol}\right\rangle=0\, .
\label{centralmass}
\end{equation}
The general consideration in Sec.~\ref{ssec:local} proves this to  all
orders. It is important that in proving the fermion-boson cancellation
no integration by parts is made. Thus, the statement of cancellation
is insensitive to the particular choice of the boundary conditions.
(As in Sec.~\ref{ssec:local}, they must be ${\cal N}=1$
supersymmetric,
of course.)

So far everything looks very similar to the minimal model treated in
Secs.\ \ref{sec:minimal} and \ref{regularization}. The
only distinction is the complex structure inherent
to the theory with four supercharges. This structure is essential, it is
just this complexification that leads to the nonrenormalization
theorem for the superpotential and for the soliton mass.

The relation~(\ref{centralmass}) makes finding the one-loop
quantum correction a
simple job, similar to the one performed in Sec.~\ref{sec:onemass}. A
somewhat
novel feature is
that the quantum fluctuations of $\chi_2$ at $z\to \infty$ and $z\to
-\infty$ are different,
\begin{equation}
\langle \, \chi_2^2 \, (z\to \pm\infty ) \rangle= \frac{1}{4\pi} \ln
\frac{4M_r^2}{m^2 (q\pm p)^2}\;.
\end{equation}
The result for the soliton mass at one loop is
\begin{equation}
M=\frac{1}{6}\,\frac{m^3}{\lambda^2} - (1-q)\,\frac{m}{4\pi} \left[\ln
\frac{4M_r^2}{m^2} +2\right] -\frac{m}{8\pi}\left[ q\ln \,(q^2 -p^2)^2
+p
\ln
\frac{(q+p)^2}{(q-p)^2}\right]
\,,
\end{equation}

The generalization to the case of arbitrary ${\cal W}(\varphi_1,
\varphi_2)$ presents no problem,
\begin{equation}
M=\left\{ {\cal W}+\sum_{k=1,2}\frac{{\cal W}_{kk}}{8\pi}\left[\ln
\frac{4M_r^2}{{\cal W}_{kk}^2} +2\right]\right\}_{\{ \varphi\}=
\{ \phi_0 (z=+\infty)\}} \!\!\!- \left\{ \{ \varphi\}=\{\phi_0
(z=-\infty)\}\right\}\,,
\end{equation}
where
$$
{\cal W}_{kl}=\frac{\partial^2 {\cal W}}{\partial \varphi_k \partial
\varphi_l}\,,
$$
and $\{\phi_0 (z)\}$ denotes the kink solution, $\varphi_1 (z)=\phi_0
(z)$,
$\varphi_2 (z)=0$.
We see that there is no correction in  the
case of unbroken
${\cal N}=2$ supersymmetry, $\sum {\cal W}_{kk}=0$,
as it should be due to nonrenormalization of the superpotential.

In the case of the soft breaking by the mass term,
\begin{equation}
{\cal W}={\cal W}_{\rm harmonic} +\frac{1}{2}\, \mu \,\phi_2^2\,,
\qquad {\cal W}_{22}=-{\cal W}_{11} +\mu\,,
\end{equation}
there is no ultraviolet divergence in the correction to the
soliton mass. Moreover,  the anomaly, presented by the
term 2 in the square brackets, also goes away.
Indeed,  these contributions are proportional to ${\cal W}_{11}+ {\cal
W}_{22}=\mu$ and are the same at both boundaries $z=\pm\infty$.

It means that we can treat the field
$\Phi_2$ as the ultraviolet regulator for $\Phi_1$, 
provided the $\Phi_2$ mass is much larger than that 
of $\Phi_1$, i.e. $\mu\gg m_1$.
In this limit we get
\begin{equation}
M=\left\{ {\cal W}+\frac{{\cal W}_{11}}{8\pi}\left[\ln
\frac{\mu^2}{{\cal W}_{11}^2} +2\right]\right\}_{\varphi_1=
\phi_0 (z=+\infty)} \!\!\!- \left\{ \varphi_1=\phi_0
(z=-\infty)\right\}\,.
\end{equation}
In this way we reproduce again the result~(\ref{massqone}) for the
soliton mass including the anomalous part (up to the substitution
$2M_r\to
\mu$).

\section{Discussion}
\label{sec:disc}
\setcounter{equation}{0}

The residual supersymmetry in the soliton background field
entails a mode degeneracy \cite{DADI}, in much the same way as
in the four-dimensional problem of instantons. This degeneracy
manifests itself in the complete cancellation of the quantum
corrections, order by order in the (small) coupling constant.
In the soliton problem this cancellation takes place in the
quantity $\langle {\rm sol}| H-{\cal Z}|{\rm sol}\rangle$. The soliton
mass  coincides with the central charge to all orders. This property,
once established at the classical level, cannot disappear at the level
of quantum corrections, at least  in the weak coupling regime.

Surprisingly, this question has caused an ongoing  debate in the
literature. In many works, including some recent ones, it is asserted
that  BPS saturation is lost at the one-loop level. Our conclusions are
categorically against this point of view. The reason explaining the
discrepancy between the (independently calculated in the literature)
soliton mass and central charge is the omission of the anomalous term
in ${\cal Z}$. We have established the existence of the anomaly
(\ref{ccdanoc}) by four independent methods. Although our explicit
calculations were carried out in one loop, we argue that Eq.
(\ref{ccdanoc}) is valid in the operator form to all orders.  In both
models considered, SPM and SSG, the contribution of the anomaly in
the central charge is negative, it is finite (i.e. contains no ultraviolet
logarithms) and makes ${\cal Z}$ smaller.  Without the anomalous
contribution, the would-be value of
${\cal Z}$ apparently  exceeds $M_{\rm sol}$ (e.g. \cite{NSNR}), in
sharp contradiction with the general principles. In fact, Ref.
\cite{NSNR} carries a special section intended to reconcile the apparent
contradiction between the general requirement that $M_{\rm sol}
\geq {\cal Z}$ and the concrete result of \cite{NSNR} according to which
$M_{\rm sol}< {\cal Z}$. Needless to say that the contradiction is
automatically eliminated once the anomaly is taken into account.

To the best of our knowledge, the anomaly in the central charge in
two-dimensional models with  minimal supersymmetry
was not discussed previously. However, indirectly its existence
could have been inferred from well-known results. For instance, the
inspection of the exact $S$ matrix in the SSG model (e.g. \cite{Ahn})
shows that the parameter $v^2$ in the superpotential (see Eq.
(\ref{spotsg})) and an analogous parameter in the on-shell scattering
amplitudes are {\em not} identical, the latter is different from the
former by a finite renormalization. In our language this
 finite renormalization can be traced back to the anomaly.

Much of the controversy in the literature is focussed around  the
boundary conditions. It is not uncommon to choose them in a
non-supersymmetric manner, e.g. \cite{KR,RN,NSNR}. It is quite obvious
that \mbox{Eq.\ (17)} in~\cite{NSNR} is incompatible with the residual
supersymmetry, see \mbox{Eq.\ (\ref{halfsusytr})}. Then the manifest
cancellation of the boson and  fermion corrections is lost, and one has
to calculate separately the boson and fermion correction. We hasten to
add that such an approach {\em a priori} is fully legitimate, there is
nothing wrong with it. Moreover, the value of the one-loop soliton mass
obtained in this way  in Refs. \cite{NSNR,Jaffe} and some previous
works is perfectly correct. The non-SUSY approach to the kink mass
calculation has its advantages since the calculation is based on the
dynamics in the bulk, and the problem of untangling the ``boundary
condition contamination" essentially does not arise.  On the other hand,
it totally misses all advantages provided by supersymmetry.

The approach we follow, which exploits in full the advantages of
supersymmetry, is very economic. Nothing comes for free, however. In
the SUSY-based approach there is a subtle point.  If one considers just
the soliton mass {\em per se}, i.e. the total energy of the system in the
box,  the boundary conditions may have a drastic impact on the result.
They distort  the soliton state under consideration in the domain
adjacent to the boundary, see Sec.\ \ref{ssec:separation}. For instance,
if the boundary conditions on
$\chi$ at $z=\pm L/2$ are periodic, they force $\langle\chi (\pm L/2)
\rangle$ to be the same while in the undistorted soliton
$\langle\chi (L/2)\rangle$ and $\langle\chi (-L/2)\rangle$
have different signs. This distortion happens even with the
supersymmetric boundary conditions. For instance, in \cite{CM} the
following condition on the derivative of
$\chi$ was suggested: $(\partial_z + F')\chi (L/2) = - (\partial_z +
F')\chi (-L/2)$, along with the antiperiodic boundary condition on
$\chi$, compatible  with (\ref{halfsusytr}). In this case the boundary
conditions enforce $\langle\bar\psi \psi\rangle_{ L/2}= \langle\bar\psi
\psi\rangle_{-L/2}$, while for the undistorted soliton
$\langle\bar\psi\psi\rangle_{L/2}$ must be equal to $-\langle\bar\psi
\psi\rangle_{ -L/2}$.

It is easy to see that the contamination of the soliton near the
boundaries adds to its mass a quantity of the same
order of magnitude as the one-loop correction. Therefore, it is not so
surprising that a spectrum of predictions for the corrected soliton
mass is obtained. These predictions refer to the contaminated soliton,
rather than the genuine ground-state soliton. It is even conceivable
to design a state whose  mass at one loop is exactly equal to the
classical soliton mass. To this end one requires $\chi$ and $\psi_2$
to vanish at the boundary. This is a possible boundary condition,
consistent with (\ref{halfsusytr}). This sensitivity to the boundary
conditions is due to the fact that the soliton mass is expressed, {\em
via} SUSY, in terms of the central charge, a quantity defined
at the boundary.

It is clear that the problem with the near-boundary domain is totally
unphysical. The effects due to  the finite-size
box must be cut off and discarded.
The physical ground-state soliton is certainly  insensitive to
the conditions we impose infinitely far away from its center.
To solve  this problem it is advantageous to consider the local form of
BPS saturation condition, see \mbox{Eq.\ (\ref{locenergy})}. Then,
by  inspecting the (one-loop corrected) soliton profile, or the energy
distribution, we can easily figure out which contribution to the mass
comes from the undistorted soliton and which is related to
distortions near the boundary.  The latter must not be included.
This approach is close to that suggested in \cite{HY} long ago, the
work which was in essence  ignored in the later publications.
Conceptually, we do the same; certainly, we do include the anomaly
omitted in Ref. \cite{HY}.
We expand this approach in various directions, e.g. we consider the
soliton profile in the minimal (one-superfield) model, and discuss a
two-superfield model presenting a deformation of the extended
${\cal N}=2$ model.

If one is not interested in the soliton profile, one can calculate the
soliton mass {\em per se}  from the central charge, by shifting the
calculation  of the latter a little bit, from the boundary of the box,
inside the fiducial domain. The edge of the fiducial domain is still
(infinitely) far from the soliton center, and, hence,  the central
charge is determined by  the flat vacuum calculation.

In the ${\cal N}=2$ models the classical expression for the soliton
mass stays intact -- there are no quantum corrections.
The corresponding nonrenormalization theorem is in essence
identical to that valid for the four-dimensional domain walls
\cite{DS,CS}. The ${\cal N}=2$ SUSY can be broken softly down to
${\cal N}=1$ by a mass term of one of the superfields.
The one-loop correction to the central charge then is finite,
since the mass term plays the role of the ultraviolet regulator.
The expression for the soliton mass is in full accord
with the general relation (\ref{ccdanoc}).

One last remark concerning the recent publications [14,15].
To an extent, our study was  inspired by the thorough analysis of Ref. [14],
which clearly exhibited the problem: the kink mass received a correction while
the central charge stayed intact. In our view, the calculations presented in [14]
are correct. Our explanation of the  results obtained  in [14] is that they refer  to
the Hamiltonian {\em corrected} by the anomaly and to the {\em uncorrected}
central charge.  These quantities, however, are not the ones related by
supersymmetry. The authors of the recent paper [15] claim a double success: the
correct value of the kink mass as well as the BPS saturation. Their calculation of the
kink mass is quite instructive. In particular, it indirectly hints at the anomaly  in
the form of the relation $k\tan (\delta_B -\delta_F)= m$ between
the boson  ($\delta_B$) and fermion ($\delta_F$) phase shifts. As for the central
charge, it was fixed in [15] to be equal to the kink mass by supersymmetry, the
assertion ascending to Ref.\ [4].  The topological nature of the charge was not
considered. In this way the authors avoided  addressing the problem  of the
difference between the  uncorrected and corrected forms of the central charge.
It was  implied that renormalization fixes the central charge uniquely -- this is  the
point we cannot accept.

The BPS saturation, as we have demonstrated, is a direct consequence
of the surviving supersymmetry. It cannot be broken as long as this
symmetry is maintained. The presence of the anomalous term in the
central charge fully resolves the long-standing riddle.  

\section*{Acknowledgments}
\label{Acknowledgments}
\addcontentsline{toc}{section}{\numberline{}Acknowledgments}

M.S. and A.V. would like to thank  I. Kogan  for a valuable contribution
at an  initial stage of this work. Discussions with B.~Chibisov,
R.L.~Jaffe,
A.~Larkin, A.~Marshakov and A.~Ritz are gratefully acknowledged.

This work was supported in part by DOE under the grant number
DE-FG02-94ER40823.

\newpage

\addcontentsline{toc}{section}{\numberline{}References}


\begin{thebibliography}{99}

\bibitem{WO}
 E. Witten and D. Olive, {\it Phys. Lett. } {\bf B78} (1978) 97.

\bibitem{DADI}
A. D'Adda and P. Di Vecchia, {\it Phys. Lett.} {\bf B73} (1978) 162.

\bibitem{KR}
R.K. Kaul and R. Rajaraman, {\it Phys. Lett.} {\bf B131} (1983) 357.

\bibitem{IM}
C. Imbimbo and S. Mukhi, {\it Nucl. Phys.} {\bf B247} (1984) 471.

\bibitem{AHV}
A. D'Adda, R. Horsley,  and P. Di Vecchia, {\it Phys. Lett.} {\bf B76}
(1978) 298; R. Horsley, {\it Nucl. Phys. } {\bf B151} (1979) 399.

\bibitem{JFS}
J.F. Schonfeld, {\it Nucl. Phys. } {\bf B161} (1979) 125.

\bibitem{SR}
S. Rouhani,
{\it Nucl. Phys. } {\bf B182} (1981) 462.

\bibitem{AU}
A. Uchiyama, {\it Nucl. Phys. } {\bf B244} (1984) 57;  {\it Prog.
Theoret. Phys.} {\bf 75} (1986) 1214.

\bibitem{AU1}
A. Uchiyama, {\it Nucl. Phys. } {\bf B278}
(1986) 121.

\bibitem{HY}
H. Yamagishi, {\it Phys. Lett.} {\bf B147} (1984) 425.

\bibitem{CM1}
A.K. Chatterjee and P. Majumdar,  {\it Phys. Lett.} {\bf B159} (1985)
37.

\bibitem{CM}
A.K. Chatterjee and P. Majumdar, {\it Phys. Rev.} {\bf D30} (1984)
844.

\bibitem{RN}
A. Rebhan and P. Nieuwenhuizen, {\it Nucl. Phys. } {\bf B508} (1997)
449.

\bibitem{NSNR}
H. Nastase, M. Stephanov, P. Nieuwenhuizen, and A. Rebhan,
{\it Topological boundary conditions, the BPS bound, and elimination
of ambiguities in the quantum mass of solitons},
hep-th/9802074; {\it Nucl. Phys. } {\bf B}, to appear.

\bibitem{Jaffe}
N. Graham and R.L. Jaffe,
{\it Energy, Central Charge, and the BPS Bound for 1+1 Dimensional
Supersymmetric Solitons}, hep-th/9808140.

\bibitem{NSVZbeta}
V. Novikov, M. Shifman, A. Vainshtein and V. Zakharov,
{\it Nucl. Phys. } {\bf B229} (1983) 381;
{\it Phys. Lett.} {\bf B166} (1986) 329.

\bibitem{DS}
G. Dvali and M. Shifman, {\it Nucl. Phys.} {\bf B504} (1997) 127.

\bibitem{CS}
B. Chibisov and  M. Shifman, {\it Phys. Rev. } {\bf D56} (1997) 7990.

\bibitem{WZ}
J. Wess and B. Zumino,  {\it Phys. Lett.} {\bf B49} (1974) 52.

\bibitem{GRS}
J. Iliopoulos and B. Zumino, {\it Nucl. Phys.} {\bf B76} (1974) 310;
P. West, {\it Nucl. Phys.} {\bf B106} (1976) 219;
M. Grisaru, W. Siegel, and M. Ro\v{c}ek,
{\it Nucl. Phys.} {\bf B159} (1979) 429.

\bibitem{mv}
M.B. Voloshin, {\it Phys. Rev.} {\bf D47} (1993) 357.

\bibitem{bs}
B.H. Smith, {\it Phys. Rev.} {\bf D47} (1993) 3518.

\bibitem{MiS}
M. Shifman, {\it Phys. Reports} {\bf 209} (1991) 341.

\bibitem{NSVVZ}
V.~Novikov {\it et al.}, {\it Nucl. Phys.} {\bf B229} (1983) 394.

\bibitem{EW}
E. Witten, {\it Nucl. Phys. } {\bf B202} (1982) 253.

\bibitem{MSAV}
M. Shifman and A. Vainshtein, {\it Nucl. Phys.} {\bf B277} (1986)
456.

\bibitem{gk}
L. Gendenshtein and I. Krive,
{\it Usp. Fiz. Nauk} {\bf 146} (1985) 553 [{\it
Sov. Phys. Uspekhi} {\bf 28} (1985) 645].


\bibitem{schwabl}
F. Schwabl, {\em Quantum Mechanics}, Springer-Verlag, Berlin, 1988.

\bibitem{rajaraman}
R. Rajaraman, {\it Solitons and Instantons}, North-Holland,
Amsterdam,
1982.

\bibitem{mv1}
M.B. Voloshin, {\it Phys. Rev.} {\bf D47} (1993) 2573.

\bibitem{brown}
L.S. Brown, {\it Phys. Rev.} {\bf D46} (1992) 4125.

\bibitem{Ahn}
R.~Shankar and E.~Witten, {\it Phys. Rev.} {\bf D17} (1978) 2134;\\
A.M.~Tsvelik, {\it Yad. Fiz.} {\bf 47} (1988) 272 [{\it Sov. J. Nucl. Phys.}  {\bf 47}
(1988) 172];\\
P. Fendley and K. Intriligator, {\it Nucl. Phys.} {\bf B372} (1990) 533;\\
  C. Ahn, D. Bernard and A. LeClair, {\it Nucl. Phys.} {\bf B346} (1990)
409; \\C. Ahn, {\it Nucl. Phys.} {\bf B354} (1991) 57.

\bibitem{Vafa}
S.~Cecotti and C.~Vafa, {\it Commun. Math. Phys.} {\bf158} (1993) 569.

\bibitem{Jackiw}
R. Jackiw and C. Rebbi, {\it Phys. Rev. } {\bf D13} (1976) 3398.

\end{thebibliography}
\end{document}